\documentclass[aps,prb,showpacs,twocolumn,superscriptaddress,nobibnotes]{revtex4-1}

\usepackage{caption}
\usepackage{wrapfig}

\usepackage[utf8]{inputenc}
\usepackage[margin=1in]{geometry}
\usepackage{amsmath}
\usepackage{braket}
\usepackage{graphicx}
\usepackage{float}
\usepackage{comment}
\usepackage{float}
\usepackage{wrapfig}
\usepackage{subcaption} 
\usepackage{chngcntr}

\usepackage{color}
\usepackage[dvipsnames, table]{xcolor}
\usepackage{algpseudocode,algorithm,algorithmicx}

\begin{document}

\title{Finite and Infinite Matrix Product States \\for Gutzwiller Projected Mean-Field Wavefunctions}
\def\urbana{
	Institute for Condensed Matter Theory and IQUIST and Department of Physics, 
	University of Illinois at Urbana-Champaign, Urbana, IL 61801, USA
}

\def\caltech{
	Division of Chemistry and Chemical Engineering,
California Institute of Technology, Pasadena, California 91125, USA
}
 
\author{Gabriel Petrica}
	\email{petrica2@illinois.edu}
	\affiliation{\urbana}

\author{Bo-Xiao Zheng}
	\affiliation{\caltech}
\affiliation{AxiomQuant Investment Management LLC, Shanghai 200120, China}
 
\author{Garnet Kin-Lic Chan}
	\affiliation{\caltech}

\author{Bryan K. Clark}
	\email{bkclark@illinois.edu}
	\affiliation{\urbana}

\begin{abstract}
Matrix product states (MPS) and `dressed' ground states of quadratic mean fields (e.g. Gutzwiller projected Slater Determinants) are both important classes of variational wave-functions.  This latter class has played important roles in understanding superconductivity and quantum spin-liquids.  We present a novel method to obtain both the finite and infinite MPS (iMPS) representation of the ground state of an arbitrary fermionic quadratic mean-field Hamiltonian, (which in the simplest case is a Slater determinant and in the most general case is a Pfaffian). We also show how to represent products of such states (e.g. determinants times Pfaffians). From this representation one can project to single occupancy and evaluate the entanglement spectra after Gutzwiller projection.   We then obtain the MPS and iMPS representation of Gutzwiller projected mean-field states that arise from the variational slave-fermion approach to the $S=1$ Bilinear-Biquadratic (BLBQ) quantum spin chain.
To accomplish this, we develop an approach to orthogonalize degenerate iMPS to find all the states in the degenerate ground-state manifold.  We find the energies of the MPS and iMPS states match the variational energies closely indicating the method is accurate and there is minimal loss due to truncation error. We then present the first exploration of the entanglement spectra of projected slave-fermion states exploring their qualitative features and finding good qualitative agreement with the respective exact ground state spectra found from DMRG.  

\end{abstract}

\maketitle

\section{\label{sec:level1}Introduction}
Variational wave-functions are frequently used to understand quantum many-body systems.  Two important classes of variational wave-functions are dressed Slater determinants and tensor networks.  Dressed Slater determinants introduce correlation on top of a mean-field ground state. On the other hand,  a tensor network is represented as a network of connected tensors providing a natural framework in which to understand and represent low-entangled quantum states (see fig.~\ref{fig:mps_tensors}). 

Slater determinants (SDs) (and other generalized quadratic ground states such as Bogoliubov-de Gennes (BdG) \cite{leggett2006quantum} and Pfaffian states \cite{read2000paired}) have played a key role in the understanding of physical systems ranging from their use as the Hartree-Fock solution in quantum chemistry to being applied as a starting mean-field ansatz for strongly-correlated systems. These latter ansatz are then dressed in various ways:  Slater-Jastrow wave-functions are the de-facto standard for simulating material systems in quantum Monte Carlo; many prototypical quantum Hall states are represented as powers or products of Slater Determinants and Pfaffians; and projected mean-field states are an important starting point for probing the physics of high temperature superconductivity as well as quantum spin-liquids.  

While dressed mean-field states are often easy to represent in variational Monte Carlo (VMC), it is also often difficult to extract certain properties from VMC. Foremost among these is the entanglement spectra which is an important metric used for understanding topological phases of matter.  Even properties which can be extracted easily, such as the energy, can be statistically noisy making aspects such as optimization difficult.  Moreover, evaluating dressed mean-field states in Monte Carlo scales cubically with the system size making the approach to the thermodynamic limit costly.   Matrix product states avoid many of these problems in one dimensional systems and ladders:  they are ideally suited for extracting entanglement spectra,  computing observables exactly without any statistical noise, and directly representing (gapped) physical systems in the thermodynamic limit. 

Our main contribution in this paper is to describe a series of efficient and highly parallel algorithms which take (projected) mean field (i.e. quadratic) eigenstates and generate both finite (fMPS) and infinite (iMPS) matrix product states from them.  We will also show how to generate fMPS and iMPS for products of mean-field wave-functions.  We will then apply our approach to compute the MPS and entanglement spectra of a series of projected slave-fermion wave-functions of the bilinear biquadratic model. This example will bring to light a number of interesting aspects of generating multiple degenerate ground states from Gutzwiller projected slave-fermion systems in iMPS. 

Beyond this particular application, being able to generate a MPS from a projected SD is generically useful.  It allows for more faithful comparisons between slave-fermion  and DMRG results which often disagree on the underlying phase of spin-liquids.  It could be used to initialize DMRG with a good initial mean-field guess for certain Hamiltonians.  This can be useful both for calculations on discrete lattices as well as DMRG in the continuum.   Because there exist algorithms which build MPS on quantum computers, it immediately gives an additional approach to generate a dressed quadratic mean-field state on a quantum device. 

We are aware of two other algorithms which convert Slater determinants to MPS \cite{fishman2015compression}, \cite{wu2020tensor}. Both these are based on the idea of applying quantum gates or matrix-product operators to a simpler quantum state to generate the MPS.  Our approach differs from these techniques in two key ways: (1) we can generate the infinite MPS for a family of slater Determinants and (2)  we generate our (i)MPS by directly generating the MPS coefficients without the application of any operators to the system. We also note that ref.~\onlinecite{schuch2019matrix} represents Slater Determinants in a MPS-like framework in a Gaussian fermionic representation.  

In section II, we will describe our key algorithm for turning a SD into a (i)MPS.  In section III, we will show a series of examples for how to use this basic procedure for generating more complicated mean-field states (i.e. pfaffians) as well as states which are products of mean-field states.  Finally, in section IV we focus on computing (i)MPS for the slave-fermion states of the billinear-biquadratic model showing their entanglement spectra and energy. 

\section{Slater Determinants to MPS}\label{sec:SDMPS}

In this section we are going to show how to generate either a finite matrix-product state (fMPStoSD) or an infinite matrix product state (iMPStoSD) from a Slater determinant (SD).  This will not only be useful in its own right but will be the key operation used in the rest of this work to produce MPS for both more complicated quadratic mean field states as well as dressed versions of these states.

fMPStoSD generates the matrix product state site by site in an approach that is highly reminiscent of the site-decimation canonical technique to convert a generic wave-function (i.e. a multi-site tensor)  into a matrix product state \cite{schollwock2011density}. The typical site-decimation procedure involves performing SVD's over matrices generated by collecting different subsets of indices into the two matrix dimensions.  This general approach will become efficient to use with Slater determinants because SVD's of Slater determinants are efficient and generate sums of products of Slater determinants.

In fMPStoSD we perform a series of Schmidt decompositions over all bipartitions of our system.  Each Schmidt decomposition generates a set of Schmidt vectors; each such Schmidt vector is a Slater determinant. The MPS is then generated by taking  overlaps of these Slater determinant Schmidt vectors with each other in the correct way.

In iMPStoSD we can easily generate the bulk uniform iMPS tensor from \textit{just} two Schmidt decompositions: one for each of two ground state Slater determinants  of the same Hamiltonian defined on sufficiently large systems that differ in size by one unit cell.  Again, these Schmidt decompositions will have Slater determinant Schmidt eigenvectors. After we appropriately fix the gauge of the two Schmidt decompositions, the uniform bulk MPS tensor will be generated from appropriate overlaps of these Schmidt eigenvectors. 

\subsection{Slater Determinant $\rightarrow$  Finite MPS}
In this section, we show in detail how to convert a Slater determinant into a finite matrix product state. The Schmidt decomposition of a  Slater determinant $|SD\rangle$ on $N$ sites bi-partitioned into two regions cut between sites $i$ and $i+1$ will be notated as 
\begin{equation}
\ket{SD} = \sum_{\alpha} \lambda^{i;N}_{\alpha} \ket{L_{\alpha}^{i;N}}         \ket{R_{\alpha}^{i;N}}
\end{equation}
where $|L_\alpha\rangle$ and $|R_\alpha\rangle$ are the $\alpha$'th left and right Schmidt vectors respectively (with support in their respective subsystem) and $\lambda_{\alpha}$ is the $\alpha$'th Schmidt eigenvalue.   Note that, for a Slater determinant, each of the individual left and right Schmidt vectors are also Slater determinants and efficiently computable   \cite{cheong2004many,klich2006lower,knizia2012density,peschel2012special}.   Slater determinants are specified by a set of single-particle orbitals and all the Slater determinants in the set of right Schmidt vectors  $\{|R_{\alpha}\rangle\}$ are specified by subsets of single particle orbitals from a set of (at most) $N$ single-particle orbitals $\{\phi^R_1 ... \phi^R_N \}$ defined on the (inclusive) sites $[(i+1), \ldots, N]$.  There are, at most, $2^N$  such subsets.    Analogous statements hold for the left Schmidt vectors. 
\begin{figure}[!h]
 \begin{subfigure}{0.5\textwidth}
\includegraphics[width=0.9\linewidth]{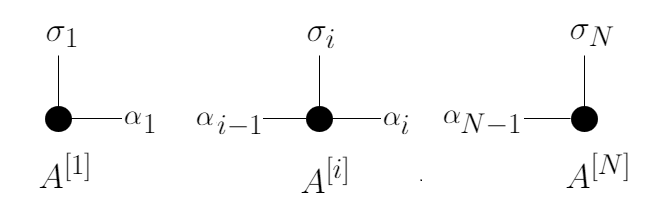} 
\caption{}
\label{fig:subim1}
\end{subfigure}
\begin{subfigure}{0.5\textwidth}
\includegraphics[width=0.9\linewidth]{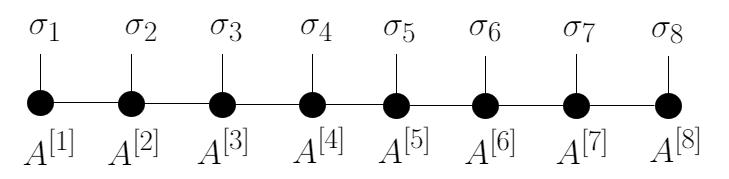}
\caption{}
\label{fig:subim2}
\end{subfigure}
\caption{a) Graphical representation of the left-boundary, bulk and right boundary $A$ tensors forming an open-boundary matrix product state. The $\alpha$'s are the virtual indices; the $\sigma$'s are the physical indices. b) Graphical representation of an $8$-site matrix product state}
\end{figure}\label{fig:mps_tensors}

A general matrix product state can be written as
\begin{equation}
\ket{MPS} = \sum_{\{\sigma\},\{a\}}
A^{[1] \sigma_1}_{1,a_1}\cdots A^{[N] \sigma_N}_{a_{N-1}, 1}
\ket{\sigma_1 \cdots \sigma_N}
\end{equation}
\noindent
where $A^{[k]}$ is the $k$'th three-tensor specified by the physical index $\sigma_k$ (e.g. occupancy or spin) and the virtual indices $\left(\alpha_{k-1},\alpha_k\right)$ \cite{schollwock2011density}.  
To generate the MPS of a Slater determinant, we compute each three-tensor $A^{[i+1]}$ as
\begin{equation}\label{eqn:ten}
A^{[i+1] \sigma_{i+1}}_{\alpha_{k}\alpha_{k+1}} = \left( \bra{\sigma_{i+1}}  \otimes \bra{R^{i+1;N}_{\alpha_{k+1}}} \right) \ket{R^{i;N}_{\alpha_k}}
\end{equation}
giving a matrix which is in right canonical form, i.e. $\sum_{\sigma} A^{[i+1]\sigma}\left(A^{[i+1]\sigma}\right)^\dagger=I$. Note that this procedure is very similar to the one which transforms a vector into a MPS \cite{schollwock2011density} and works for the same reason: the sets $\{|R^{i;N}_{\alpha}\rangle\}$ and $\{|\sigma_i\rangle \otimes |R^{i+1;N}_{\beta}\rangle \}$ span the same space and therefore there is a transformation $A$ which rotates between them.  In practice, we keep the bond-dimension of $A$ controlled by only computing the Schmidt vectors whose Schmidt values are above a certain threshold $\epsilon$.  
This can be done without computing any Schmidt eigenvector with eigenvalue less than $\epsilon$. 
Here we have focused on the bulk tensors and slight modifications need to be made for the boundary tensors $A^{[1]\sigma_1}$ and $A^{[N]\sigma_N}$ (see supplement ~\ref{supplement:boundary_tensors}).
\begin{figure}[!h]
\includegraphics[width=\columnwidth]{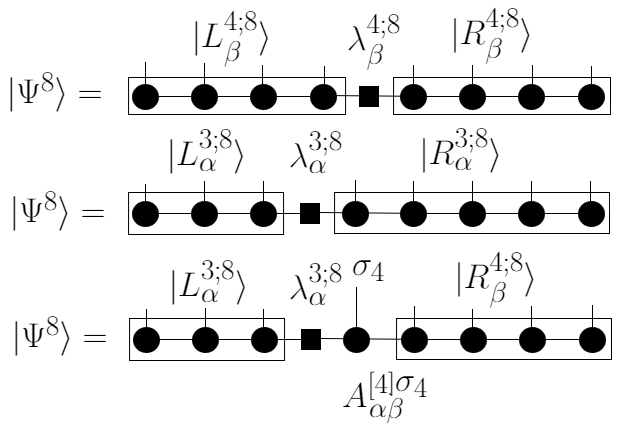} 
\caption{ \textbf{Top: }Graphical representation of the Schmidt decomposition of the wavefunction $\ket{\Psi^8} = \sum_\beta \lambda_{\beta}^{4;8} \ket{L_\beta^{4;8}}\ket{R_{\beta}^{4;8}}$ over a bipartition $[1,\ldots,4]\times[5, \ldots, 8]$ of an $8$-site system.  \textbf{Middle, Bottom}: Two additional ways of representing the quantum state $|\Psi_8\rangle$. The tensor $A^{[4]}$ is constructed by having the overlap of the right five sites of the bottom two figures equal one. \label{fig:subim1} }
\end{figure}
We now describe how to efficiently evaluate the matrix elements of each $A$. We start by noting that 
$| \sigma_{i+1}|\rangle \otimes |R^{i+1;N}\rangle$ is also a Slater determinant.  It is specified by the single particle orbitals  
\begin{equation}
     \{[0\phi_a],[0\phi_b],\cdots, [0\phi_c], \phi^{i+1}\}
\end{equation}
where $[0\phi_a]$ is the single particle orbital with coefficients in the lattice basis
$[0,\phi_a(i+2),\phi_a(i+3),\cdots,\phi_a(N)]$ and 
$\phi^{i+1}$ is the single-particle orbital in analogous notation,  $[1,0,\cdots,0]$. Eqn. \ref{eqn:ten} then reduces to the overlap of two Slater determinants of size $(N-i) \times (N-i)$ which can be computed in $O(N^3)$ time.  

While naively each element of $A$ requires such a computation, there is a significant overlap in these different computations which reduce the naive compuational complexity of the tensor computation. There are two steps in computing the overlap of two Slater determinants:  evaluating the overlap matrix between all pairs of single particle orbitals that make up the two determinants and computing the determinant of this overlap matrix.  All the Slater determinants used in the ket (respectively bra) of eqn. \ref{eqn:ten} (over different terms in A) come from  a subset of single-particle orbitals of the N-orbital set  $\{\phi_1^R,..,\phi_N^R\}$. We can compute the overlap matrix of all these respective single-particle orbitals once per three-tensor $A$ at a cost of $O(N^3).$  The entries of $A$ are then determinants of submatrices of this overlap matrix.   While naively each determinant also costs $O(N^3)$ to compute, the submatrices differ only in the bottom $\log_2 D$ columns and right $\log_2 D$ rows where $D$ is the bond-dimension of $A$; determinant update formulas can then be used to accelerate this computation letting each determinant be computed in time $O(N^2 \log_2 D)$ after an initial $O(N^3)$ operation to evaluate the inverse of the upper-left $(N-\log_2 D) \times (N-\log_2 D)$ block of the overlap matrix.   The whole evaluation of each tensor $A$ can be done in $O(N^3) + O(D^2 N^2 \log_2 D)$ time. This can be further attenuated somewhat by more aggressive use of determinant update formulas \cite{harville1998matrix}.

Notice that there are significant parts of this algorithm that can be run in parallel. Each three-tensor $A$ can be computed separately.  Within each $A, $ the Schmidt decomposition can be partially parallelized; each element of the overlap matrix can be computed in parallel; and, after the initial evaluation of the inverse of the upper-left block of the overlap matrix, each determinant can then be computed in parallel.

\subsection{Gapped Slater Determinant $\rightarrow$ Infinite MPS}
\begin{figure}[!h]
\includegraphics[width=1\linewidth]{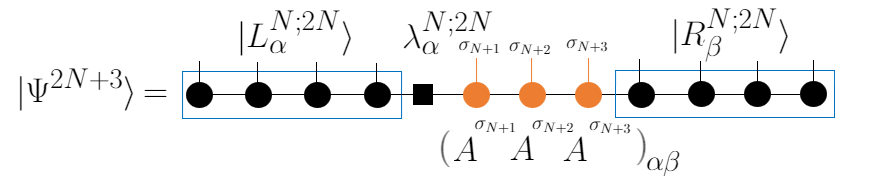} 
\caption{We obtain the finite MPS $\ket{\psi^{2N+n}}$ by inserting $n$ (in the figure $n=3$) $A$ iMPS bulk tensors between the left $\ket{L^{N/2;N}}$ and right $\ket{R^{N/2;N}}$ Schmidt vectors obtained from $\ket{\psi^{2N}}$. 
}
\label{fig:wfn_pred}
\end{figure}

The above procedure generates a finite MPS approximation (the accuracy of the representation is given as an input to the algorithm) of any Slater determinant. In this section, we describe how to generate an infinite MPS from the Slater determinant ground state of a gapped mean-field Hamiltonian.  This infinite MPS can be described by left $L$ and right $R$ boundary tensors which sandwich the bulk tensor $A$ giving us an infinite matrix product state of the form,
\begin{multline}
    \ket{\textrm{iMPS}} = \sum_{\sigma} L^{\sigma_L} \ldots A^{\sigma_{n-1}}A^{\sigma_n}A^{\sigma_{n+1}}
    \ldots R^{\sigma_R}  \times \\
    \ket{\sigma_L \ldots \sigma_{n-1} \sigma_n \sigma_{n+1} \ldots \sigma_R}
\end{multline} 
with an arbitrary number of bulk tensors $A$.  $L$ and $R$ are tensors which span a fixed number $k$ of sites.   Note that any thermodynamic observable can be computed directly in the thermodynamic limit of the Slater Determinant using only the bulk tensor $A$. In addition, we can compute the amplitude for the Slater determinant on any (large enough) system size, by inserting the corresponding number of bulk tensors between the boundary tensors $L$ and $R$ (i.e. to generate the MPS for a $N$ site Slater determinant from the infinite MPS, we therefore use $N-2k$ bulk tensors $A$); see fig.\ref{fig:wfn_pred}. 

To generate the iMPS, we start off by producing two Slater determinants defined on $2N$ and $2N+1$ sites, where $N$ is sufficiently large such that the entanglement spectrum is constant over cuts in the ``bulk'' of the wave-functions.  For gapped systems, we generically expect the entanglement spectrum over the bulk to be constant; see fig. ~\ref{fig:es_bulk} in supplement ~\ref{supplement:constant_es} for an example of this.  We then generate the Schmidt decompositions
\begin{eqnarray}
    |\Psi^{2N}\rangle & =  & \sum_\alpha \lambda^{N;2N}_{\alpha} |L_\alpha^{N;2N}\rangle|R_\alpha^{N;2N}\rangle \\
   | \Psi^{2N+1}\rangle & = & \sum_\alpha \lambda^{N;2N+1}_{\alpha} |L_\alpha^{N;2N+1}\rangle|R_\alpha^{N;2N+1}\rangle 
\end{eqnarray}
Both $|L_\alpha^{N;2N}\rangle$ and $|L_\alpha^{N;2N+1}\rangle$ are going to be the same up to a gauge freedom. We fix this gauge freedom by defining a unitary
\begin{eqnarray}
    C_{\alpha \beta}^{N;2N+1}   = & 0  & \textrm{ if } \lambda_{\alpha} \neq \lambda_{\beta}\\
                    = & \bra{L^{N;2N}}_{\alpha}\ket{L^{N;2N+1}}_{\beta} & \textrm{ otherwise } \nonumber
\label{eqn:C}
\end{eqnarray}
which rotates between Schmidt eigenvectors with the same Schmidt eigenvalue allowing the state on $2N+1$ sites to be defined as 
\begin{equation}
    \ket{\Psi^{2N+1}} = \sum_{\alpha\gamma}\ket{L^{N; 2N}}_{\alpha}\lambda^{N;2N+1}_{\alpha} C^{N;2N+1}_{\alpha\gamma}\ket{R^{N;2N+1}}_{\gamma}
\end{equation}
Then the tensor $A$ for the iMPS is 
\begin{equation}
A^{\sigma_{N+1}}_{\alpha\beta} = \sum_{\gamma} C^{[2N+1]}_{\alpha\gamma}\bra{\sigma_{N+1}|\bra{R^{N;2N}}_{\beta}}
  \ket{R^{N,2N+1}}_{\gamma}
\end{equation}
\begin{figure}[!h]
\includegraphics[width=\columnwidth]{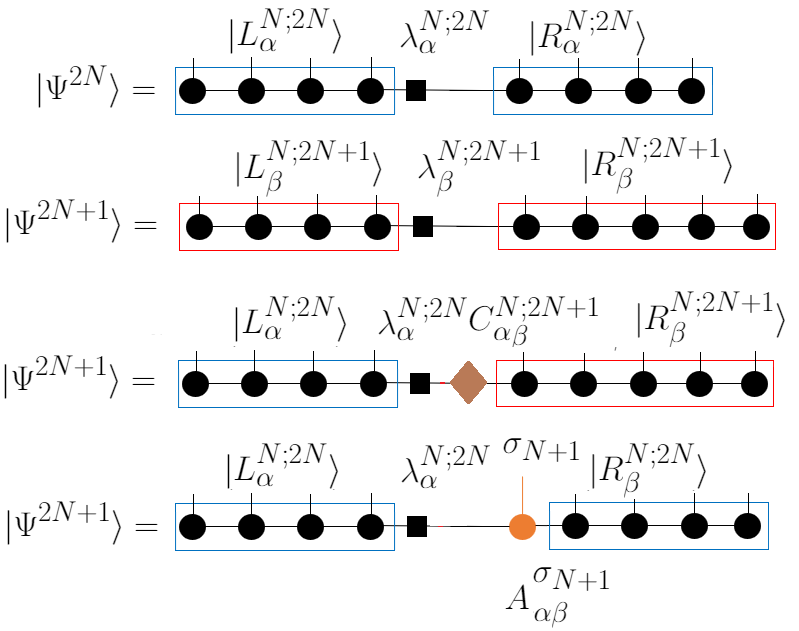} 
\caption{Illustration of the algorithm for generating the infinite MPS representation of a Slater determinant.  Lines 1 and 2 correspond to the standard Schmidt decomposition after site $N$ of wavefunctions defined on $2N$ and $2N+1$ sites.   For line three, we use our gauge freedom to replace the left-Schmidt eigenvalues  $\lambda^{N;2N+1}$ and eigenvectors $\ket{L^{N;2N+1}}$ with eigenvalues $\lambda^{N;2N}$ and eigenvectors $\ket{L^{N;2N}}$ rotated by $C$ where $C$ is defined in eqn.~\ref{eqn:C}.  Finally, the tensor $A$  is constructed by having the overlap of the right $N+1$ sites (including $C$) of the bottom two lines equal one. }
\label{fig:image2}
\end{figure}

As in the finite MPS case, we have that the single-particle orbitals of the Slater determinant $|\sigma_{N+1}\rangle |R^{N;2N}\rangle$ are shifted to the right with an initial zero as their first element.  The overlap of this tensor can be computed in exactly the same way as for the finite MPS case.  Here, though, we only need to evaluate one tensor $A$ instead of a tensor per site, with the assumption that we are using an iMPS defined by a single tensor (i.e. single site unit cell) $A$.  This process can be generalized to multi-site unit cells as well (see supplement \ref{supplement:multi-site}).  
Note that by directly applying the finite MPS algorithm to large systems to try to find the bulk tensor $A$ will fail because the gauge freedom available in the tensors will prevent a single identical bulk tensor from being produced at each step. 

\subsection{Numerical Validation}
We numerically validate our algorithms by applying fMPStoSD and iMPStoSD on the ground state of the Su-Schrieffer-Heeger(SSH) model \cite{su1979solitons},
\begin{multline}
H_{SSH} = v \sum_n \left( c^{\dagger}_{n,1} c_{n,2} +  \text{h.c.}\right)+ \\ w\sum_n \left(c^{\dagger}_{n+1,1}c_{n,2} + \text{h.c.} \right)
\label{eqn:SSH}
\end{multline}
The model describes spinless fermions on a 1D lattice, with a 2 site unit cell made up of $A$,$B$ sites, with different (real) parameters for intra-cell hopping ($v$) and intercell hopping ($w$). 
It admits two different quantum ground states, distinct in their topological properties: a trivially gapped phase for $v>w$ and a (symmetry protected) topological gapped ground state, characterized by the presence of two zero-energy edge modes inside the gap, for $v<w$,  separated by a quantum critical point at $v=w$.

\begin{figure*}
  \includegraphics[width=\textwidth]{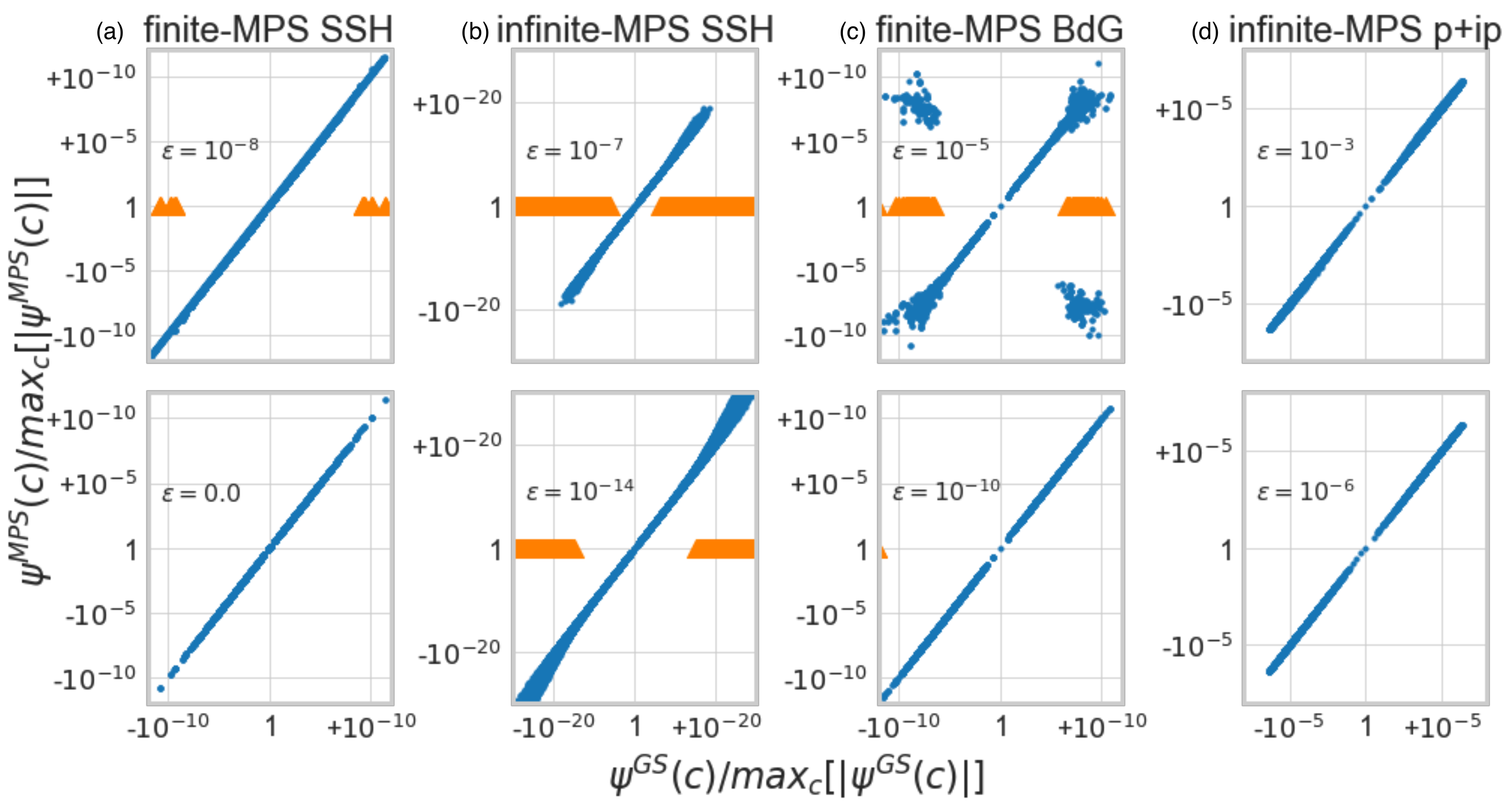}
  \caption{
  Comparison of amplitudes (normalized by the largest amplitude seen) between our fMPS/iMPS wave-functions and the exact SD.  The largest (normalized) amplitude is at the ``origin'' of the graphs with smaller amplitudes toward both edges.  Orange triangles are values at which the fMPS/iMPS gives zero amplitude; for these the "y" coordinate is arbitrarily set to 1.  The amplitudes for the top row are less accurate as they are generated with larger MPS thresholds  $\epsilon$. 
   (a) and (c) compare all amplitudes for $N=8$ on ($v=1.0$; $w=0.6$; eqn.~\ref{eqn:SSH}) and ($t=1$; $\mu=3$; $\Delta=1$; eqn.~\ref{eqn:bdg}) respectively. 
  (b) We compare 459428 random configuration (top and bottom are different configurations) between the $N=24$ Slater determinant with  ($v = 1.0$; $w=0.6$) of eqn.~\ref{eqn:SSH} and a MPS generated from 8 uniform iMPS bulk tensors (generated from SD on $N=16,17$ sites) sandwiched between the  $8$-left and $8$-right tensors from the $16$-site Slater determinant.   
  (d) We compare $49972$ (top) and $34933$ (bottom)  random configuration between the $N=32$ p+ip Pfaffian groundstate with  ($t = 1.0$; $\Delta =-1$  $\mu=-2.2$) of eqn.~\ref{eqn:hpip} and a MPS generated from 8 uniform iMPS bulk tensors (generated from SD on $N=24,25$ sites) sandwiched between the  $12$-left and $12$-right tensors from the $24$-site Pfaffian. 
  }
  \label{fig:num_valid}
\end{figure*}

We will discuss here the trivial ground state. A small subtlety related to choosing the same gapless boundary mode in the Slater determinant wave-functions used for generating the uniform tensor in the iMPS procedure is delegated to the supplement \ref{supplement:imps_edge}.  For the finite Slater Determinant, we compare the MPS we generate using two different truncation values against the exact Slater Determinant by comparing all of the amplitudes (see 1st column in fig. \ref{fig:num_valid}). For the infinite case, we generate the iMPS and then use the bulk tensor we have computed along with the boundary tensors to compute amplitudes for a much larger system and again compare amplitudes against the exact solution for that much larger system (see 2nd column in fig. \ref{fig:num_valid}). In both cases, we find that the amplitudes are in very good agreement for all amplitudes down to the Schmidt eigenvalue cutoff.

\section{General (Dressed) Quadratic Mean Fields}\label{ref:MFMPS}
In sec.~\ref{sec:SDMPS} we showed how to generate a matrix product state from a Slater determinant. In this section, we show that this machinery gives us the means to generate the matrix-product state representations of ground states of arbitrary quadratic mean-field Hamiltonians.   

All quadratic Hamiltonians can be easily diagonalized using a canonical transformation \cite{ripka1986quantum}. Without loss of generality, in our derivations, we will use translation invariant systems for ease of presentation. 
We will first go through two canonical examples. In \ref{subsec:bdgmps} we will show how to produce the MPS representation of groundstates of BdG Hamiltonians which are Slater determinants in disguise. In \ref{subsec:pffmps}, we show how to  compute the MPS representation of the p-wave pairing ground state of the Kitaev p+ip chain \cite{kitaev2001unpaired}.  We then generalize this result to general Pfaffian wave-functions which are the most general quadratic mean field ground states. Finally, we show how to take products (or powers) of quadratic mean-field Hamiltonians and turn them into (i)MPS.

\subsection{BdG $\rightarrow$ MPS }\label{subsec:bdgmps}
The key trick to convert a BdG wave-function into a MPS will be to 1) convert it to a Slater determinant through a particle hole transformation, 2) convert this Slater determinant to a MPS, and 3) then undo the particle-hole transformation in the MPS language. 

Consider a BdG Hamiltonian:
\begin{multline}
H_{BdG} = -\sum_{\langle ij \rangle, \sigma}t_{ij}\left( c^{\dagger}_{i\sigma} c_{j\sigma} + \text{h.c.}\right) - \\
\sum_{\langle ij \rangle} \Delta_{ij} \left( c^{\dagger}_{i\uparrow} c^{\dagger}_{j\downarrow} +  \text{h.c.} \right) -
\mu \sum_{i\sigma} c^{\dagger}_{i\sigma} c_{i\sigma} 
\label{eqn:bdg}
\end{multline}
Under a canonical particle-hole transformation in the $\downarrow$ sector:
\begin{equation}
    \begin{split}
        f^{\dagger}_{2i-1} = c^{\dagger}_{i\uparrow} \\
        f^{\dagger}_{2i} = c_{i\downarrow}
    \end{split}
\end{equation}
the BdG Hamiltonian becomes:
\begin{multline}\label{eqn:bdg_ph_ham}
    H^{ph}_{\text{BdG}} = - \sum_{\langle ij \rangle} t_{ij}\left( f^{\dagger}_{2i-1}f_{2j-1}  -f^{\dagger}_{2i}f_{2j} + \text{h.c.} \right) -\\
    \sum_{\langle ij \rangle} \Delta_{ij} \left( f^{\dagger}_{2i-1} f_{2j} +  \text{h.c.} \right)
    -\mu \sum_{i\sigma} \left(f^{\dagger}_{2i-1} f_{2i-1} - f^{\dagger}_{2i} f_{2i} \right)
\end{multline}
and the new vacuum is $\ket{0^{\text{ph}}} = c^{\dagger}_{1\downarrow} c^{\dagger}_{2\downarrow} \ldots c^{\dagger}_{N\downarrow}\ket{0}$, where $\ket{0}$ is the vacuum of the original theory.
The ground state of $H_{\text{BdG}}$ is thus the Slater determinant ground state of $H^{\text{ph}}_{\text{BdG}}$ on top of the new vacuum $\ket{0^{\text{ph}}}$.

Using the results from section \ref{sec:SDMPS}, we convert the  Slater determinant ground state of $H^{\text{ph}}_{\text{BdG}}$ into a MPS  $\ket{\Psi_{MPS}} = \sum_{\{\sigma\}} A^{[1]\sigma_1}\ldots A^{[2N]\sigma_{2N}} \ket{\sigma_1 \ldots \sigma_{2N}}$ where $\sigma_{2i-1}=\{0,1\}$ ($i \in [1,N])$ indicates the absence/presence of a $\uparrow$ particle and $\sigma_{2i}=\{0,1\}$ indicates the absence/presence of a \textit{hole} on top of the filled $\downarrow$ Fermi sea at site $i$.

To `undo' the particle-hole transformation, we need to deal with the fact that the $f_i$  act on the false vacuum $\ket{0^{ph}}$ (and not the real vacuum) by  swapping, for all $i$, the matrices  $A^{[2i]1}$ and $A^{[2i]0}$. Moreover, by ordering the fermionic operators by site, and then spin, the matrices $A^{[2i-1]1}$, $A^{[2i]1}$ will pick up factors of $(-1)^{i-1}$.  We can now combine these transformations giving us our final MPS for the BdG ground state of the form
$|\Phi_{GS} \rangle = \sum_{\{\sigma={0,\uparrow,\downarrow,\uparrow\downarrow\}}} B^{[1]\sigma_1}\ldots B^{[N]\sigma_N} |\sigma_1 \ldots \sigma_N\rangle $
where
\begin{align}\label{eqn:bdgmps}
B^{[i]0} &= (-1)^{i-1} \times A^{[2i-1]0} A^{[2i]1} \nonumber \\
B^{[i]\uparrow} &=  (-1)^{2i-2} \times A^{[2i-1]1}A^{[2i]1} \nonumber \\
B^{[i]\downarrow} &= (-1)^{0}\times A^{[2i-1]0}A^{[2i]0} \nonumber \\
B^{[i]\uparrow\downarrow} &= (-1)^{i-1}\times A^{[2i-1]1}A^{[2i]0}
\end{align}
This approach works both for the finite and infinite MPS as we just used our (i)MPS $\rightarrow$ Slater determinant approach as a subroutine. For the infinite MPS it produces a unit cell of size 2 as  every other $B$ differs by a sign.  

As a check of our algorithm, we consider the ground state of the BdG Hamiltonian of the form:
\begin{multline}
H_{BdG} = -\sum_{\langle ij \rangle, \sigma}t_{ij}\left( c^{\dagger}_{i\sigma} c_{j\sigma} + \text{h.c.}\right) - \\
\sum_{\langle ij \rangle} \Delta_{ij} \left( c^{\dagger}_{i\uparrow} c^{\dagger}_{j\downarrow} + c^{\dagger}_{j\uparrow}c^{\dagger}_{i\downarrow} + \text{h.c.} \right) - 
\mu \sum_{i\sigma} c^{\dagger}_{i\sigma} c_{i\sigma} 
\label{eqn:bdg}
\end{multline}
and compare the amplitudes of the exact ground state with the MPS generated (see 3rd column in fig.~\ref{fig:num_valid}).

\subsection{Pfaffian $\rightarrow$ MPS}\label{subsec:pffmps}
We will show how to generate a MPS representa-tion of the Pfaffian ground state of the Kitaev p+ip chain:
\begin{multline}\label{eqn:hpip}
    H_c = -t \sum_n \left( c^{\dagger}_{n}c_{n+1} + \text{h.c.} \right) 
    +    \Delta \sum_{n}\left( c^{\dagger}_{n}c^{\dagger}_{n+1}  + \text{h.c.} \right) \\
    +\mu \sum_n c^{\dagger}_{n}c_{n} 
\end{multline}
We first consider $H^{\text{ext}}=H_c \bigoplus H_d$ whose ground state is given by the tensor product of two identical pfaffians: 
\begin{equation}\label{eqn:analytic_c}
    \ket{GS} = \sum_{\sigma_{c},\sigma_{d}} Pf(M_{\sigma_{c}}) \ket{\sigma_{c}} \bigotimes Pf(M_{\sigma_{d}}) \ket{\sigma_{d}}
\end{equation}
where $M$ is a $N \times N$ matrix built from parameters of the model and $M_{\sigma_{c}}$ is a submatrix of $M$ obtained by selecting indices as given by the $\sigma_{c}$ configuration.

Using the local canonical transformation of fermions, 
\begin{equation}\label{eqn:local_transformation}
    \begin{split}
        c^{\dagger}_{n} =  \frac{\left(\bar{c}^{\dagger}_{n,\uparrow} + \bar{c}^{\dagger}_{n,\downarrow} \right) }{\sqrt{2}}\\
        d^{\dagger}_{n} = i\frac{\left(\bar{c}^{\dagger}_{n,\uparrow} - \bar{c}^{\dagger}_{n,\downarrow} \right) }{\sqrt{2}}
    \end{split}
\end{equation}
converts $H^{\text{ext}}$ to
\begin{multline}
    H_{BdG} = -t \sum_{n,\sigma} \left( \bar{c}^{\dagger}_{n,\sigma} \bar{c}_{n+1,\sigma} +\text{h.c.}\right) \\
    + \Delta \sum_n \left( \bar{c}^{\dagger}_{n\uparrow}\bar{c}^{\dagger}_{n+1\downarrow} +  \bar{c}^{\dagger}_{n+1,\downarrow}\bar{c}^{\dagger}_{n,\uparrow}+ \text{h.c.} \right) \\
    + \mu \sum_{n\sigma} \bar{c}^{\dagger}_{n,\sigma}\bar{c}_{n,\sigma}
\end{multline}
The transformation leaves the vacuum unchanged.
Given a BdG Hamiltonian,  we can obtain the ground-state as an MPS as done in section \ref{subsec:bdgmps}.
\begin{equation}
    \ket{GS} = \sum_{\sigma} \ldots A^{2i-1\sigma_{2i-1}}A^{2i\sigma_{2i}} \ldots \ket{\sigma_1 \sigma_2 \ldots \sigma_{2N}}
\end{equation}
where $A^{2i-1\sigma_{2i-1}}$ is the tensor on site $i$ for the $\uparrow$ local physical sector and $ A^{2i\sigma_{2i}}$ is the tensor on site $i$ for the $\downarrow$ local physical sector.

Notice that the canonical transformation given in eqn.~\ref{eqn:local_transformation} mixes the $\uparrow$, $\downarrow$ physical sectors on site $i$. Hence we can obtain the MPS for the GS in the $c,d$ space by choosing the on-site tensor in the following way:

\begin{equation}
    \ket{GS} = \sum_{\sigma} C^{1\sigma_1} C^{2\sigma_2} \ldots C^{N\sigma_N} \ket{\sigma_1 \sigma_2 \ldots \sigma_{N}}
\end{equation}
with $\sigma = \{0, c, d, cd\}$ where: 
\begin{equation}\label{eqn:mps_transformation}
\begin{split}
    C^{n,0} = A^{2n-1, 0}A^{2n, 0} \\
    C^{n,c} = \frac{A^{2n-1, 1}A^{2n, 0} + A^{2n-1, 1}A^{2n, 0}}{\sqrt{2}} \\
    C^{n,d} = i\frac{A^{2n-1, 1}A^{2n, 0} - A^{2n-1, 1}A^{2n, 0}}{\sqrt{2}} \\
    C^{n,cd} = iA^{2n-1, 1}A^{2n, 1}
\end{split}
\end{equation}
By projecting out the $d$ particles in the $\ket{GS}$ wavefunction, we obtain a Pfaffian wavefunction in the $c$-particle sector: $\ket{GS} = \text{constant} \times \sum_{\sigma_{c}} Pf(M_{\sigma_{c}}) \ket{\sigma_{c}}$.
At the level of the MPS this projection is realized by eliminating the sectors $\sigma = \{d, cd\}$. 
\begin{equation}\label{eqn:pfaffian_mps}
    \ket{\text{Pf}} = \sum_{\sigma=\{0,c\}}  C^{1\sigma_1} C^{2\sigma_2} \ldots C^{N\sigma_N} \ket{\sigma_1 \sigma_2 \ldots \sigma_{N}}
\end{equation}

\subsection{Pfaffian  $\rightarrow$ MPS Generalization}\label{sec:pffmps_general}
While we focused in the previous section on a specific example, here we consider a generic quadratic Hamiltonian $H = \sum_{n,m} C^{\dagger}_n h_{n,m} C_{m}$ with $g$ species of fermions per unit cell where the vector $C^{\dagger}_n = \left(c^{\dagger}_{n,1}, c_{n,1}, c^{\dagger}_{n,2}, c_{n,2} \ldots c^{\dagger}_{n,g},c_{n,g}\right)$. 

We form an extended Hamiltonian $H^{\text{ext}}$ which is a sum of two copies of $H$: 
\begin{multline}
    H^{ext} = \sum h^{\text{hop}}_{i\alpha,j\beta} \left(c^{\dagger}_{i,\alpha}c_{j,\beta} +d^{\dagger}_{i,\alpha}d_{j,\beta} + \text{h.c.} \right)  \\
    + \sum h^{\text{pair}}_{i\alpha,j\beta} \left(c^{\dagger}_{i,\alpha}c^{\dagger}_{j,\beta} +d^{\dagger}_{i,\alpha}d^{\dagger}_{j,\beta} + \text{h.c.} \right) 
\end{multline}

where $\alpha,\beta \in (1,\ldots, g)$. As before, its ground state $\ket{\text{GS}}^{\text{ext}} = \sum_{\sigma_{c},\sigma_{d}} Pf(M_{\sigma_{c}}) \ket{\sigma_{c}} \bigotimes Pf(M_{\sigma_{d}}) \ket{\sigma_{d}}$ is a tensor product of two identical Pfaffian wavefunctions. We then obtain the Pfaffian ground state of $H$ by projecting out all the $d$ sectors.
Under the following linear canonical transformation
\begin{equation}\label{eqn:lin_duplicate}
    \begin{split}
        \bar{c}^{\dagger}_{n,\alpha,\uparrow} = \frac{c^{\dagger}_{n,\alpha}+id^{\dagger}_{n,\alpha}}{\sqrt{2}} \\
        \bar{c}^{\dagger}_{n,\alpha,\downarrow} = \frac{c^{\dagger}_{n,\alpha}-id^{\dagger}_{n,\alpha}}{\sqrt{2}}
    \end{split}
\end{equation}
$H^{\text{ext}}$ becomes a BdG-like hamiltonian when expressed 
in terms of $c_{\uparrow}$ and $c_{\downarrow}$:
\begin{equation}\label{eqn:h_extern}
    \begin{split}
        H^{ext}_{\text{BdG}} = \sum h^{\text{hop}}_{i\alpha,j\beta} \left(\bar{c}^{\dagger}_{i,\alpha,\uparrow}\bar{c}_{j,\beta,\uparrow} +\bar{c}^{\dagger}_{i,\alpha,\downarrow}\bar{c}_{j,\beta,\downarrow} + \text{h.c.} \right) + \\
        + \sum h^{\text{pair}}_{i\alpha,j\beta} \left(\bar{c}^{\dagger}_{i,\alpha,\uparrow}\bar{c}^{\dagger}_{j,\beta,\downarrow} +\bar{c}^{\dagger}_{i,\alpha,\downarrow}\bar{c}^{\dagger}_{j,\beta,\uparrow} + \text{h.c.} \right) 
    \end{split}
\end{equation}
We can then solve for the MPS representation of the groundstate of the above BdG-like Hamiltonian using the methods described in section ~\ref{subsec:bdgmps}.

We obtain the Slater determinant ground state $\bra{\Psi_{ext}}$ by diagonalizing the particle-hole transformed $H^{\text{ext}}_{\text{BdG}}$ and then computing its MPS representation. Each unit cell $M$ is described by $2g$ tensors $A^{[M,p]\sigma}$ with $\sigma=0,1$ signifying the absence/presence of a particle of type $p \in [0,1,\ldots 2g-1]$. 
A particle of type $p=2k$ corresponds to the flavor $k,\uparrow$; a particle of type $p=2k+1$ corresponds to the flavor $k,\downarrow$.
From the above MPS (which is in $\bar{c}_{\uparrow}$, $\bar{c}_{\downarrow}$ local physical space) we construct the MPS tensors in $c$, $d$ space. In particular, the matrices describing the absence/presence of a particle of type $c_i$ on site $M$ are given by:
\begin{align}
B^{[M,i]0} & =  A^{[M,2i-1][0]}A^{[M,2i][1]}  \\
B^{[M,i]1} & = \frac{(-1)^{g(M-1)+i-1}}{\sqrt{2}} \times \nonumber \\
& \left[ A^{[M,2i-1][1]} A^{[M,2i][1]} + A^{[M,2i-1]0} A^{[M,2i][0]} \right] \nonumber
\end{align}
where $(-1)^{g(M-1)+i-1}$ takes care of fermionic ordering. 
and the MPS representation of $\ket{\Psi_{GS}}$ defined on $Ng$ sites is given by:
\begin{multline}
\ket{\Psi_{GS}} = \sum_{\{\sigma\}} \left(B^{[1,1]\sigma_{1_1}}B^{[1,2]\sigma_{1_2}} \ldots B^{[1,g]\sigma_{1_g}} \right)\ldots \times \\ 
\left(B^{[N,1]\sigma_{N_1}}B^{[N,2]\sigma_{N_2}} \ldots B^{[N,g]\sigma_{N_g}} \right) \times \\  \ket{\left(\sigma_{1_1} \sigma_{1_2} \ldots \sigma_{1_g} \right)\ldots \left(\sigma_{N_1} \sigma_{N_2} \ldots \sigma_{N_g}\right)}
\end{multline}
By suitably contracting tensors we can obtain a $N$-tensor MPS representation with physical dimension $2^g$: $\ket{Pf} = \sum_{\sigma} C^{1\sigma_1} C^{2\sigma_2} \ldots C^{N\sigma_N} \ket{\sigma_1 \sigma_2 \ldots \sigma_{N}}$. For instance, the tensor corresponding to the \textit{presence} of particles of type $t_1,t_2, \ldots t_s$ on site $M$ is 
\begin{multline}
C^{[M](t_1,t_2,\ldots,t_s)} = B^{[M,1][0]}\ldots \times \\ B^{[M,t_1][1]} B^{[M,t_1+1][0]}\ldots B^{[M,t_s][1]} \ldots B^{[M,g][0]}
\end{multline}

\subsection{Power of Slater determinants}

In this section we describe how to obtain the MPS representation of a wavefunction
\begin{equation}
\ket{\psi_{1/n}} = \sum_{r_1,r_2,\cdots r_n} \braket{r_1, r_2, \ldots r_n|\psi}^n|r_1,r_2, \ldots, r_n\rangle
\end{equation}
where $\ket{\psi}$ is a Slater determinant. We will use as an example n = 3.  Products of other mean-field wave-functions can be obtained similarly. 

We extend our N-site system to a $3N$-site system for which we label the sites as: $\{1_1 1_2 1_3 2_1 2_2 2_3 \ldots N_1 N_2 N_3\}$.  We then write a single Slater determinant (by padding and interlacing the orbitals to keep the above ordering) of the form $\ket{\psi}\bigotimes\ket{\psi}\bigotimes\ket{\psi}$ for which we then convert into an MPS  given by
\begin{multline}
\ket{MPS} = \sum_{{i_p}}
\left( B^{[1_1]i_{1_1}}B^{[1_2]i_{1_2}}B^{[1_3]i_{1_3}}\right) \ldots \times \\ \left( B^{[N_1]i_{N_1}}B^{[N_2]i_{N_2}}B^{[N_3]i_{N_3}}\right) \times \\ \ket{i_{1_1}i_{1_2}i_{1_3} \ldots   i_{N_1}i_{N_2}i_{N_3}}
\end{multline}
Projecting on the sector $i_{n_1}=i_{n_2}=i_{n_3}$ gives us the desired results
of 
\begin{equation}
\ket{\psi_{1/3}} = A^{[1]i_1}A^{[2]i_2} \ldots A^{[N]i_N} \ket{i_1 i_2 \ldots i_N}
\end{equation}
where we define 
\begin{equation}
    A^{[n]i_n} = B^{[n_1]i_{n_1}}B^{[n_2]i_{n_2}}B^{[n_3]i_{n_3}}
\end{equation}

\section{\label{sec:bibiq} Bilinear-Biquadratic S=1 model}
In this section, we use our approach to compute the MPS representation and entanglement spectra of the Gutzwiller projected slave-fermion mean-field states \cite{liu2012gutzwiller} of the  Bilinear-Biquadratic S=1 model,
\begin{equation}\label{eqn:blbq_ham}
    H = \sqrt{J_1^2 + J_2^2} \sum_{<i,j>}\left( \cos{\theta}\mathbf{S}_i \cdot \mathbf{S}_j  + \sin{\theta} \left(\mathbf{S}_i \cdot \mathbf{S}_j \right)^2\right),
\end{equation}

\begin{figure}[H]
\includegraphics[width=0.8\columnwidth]{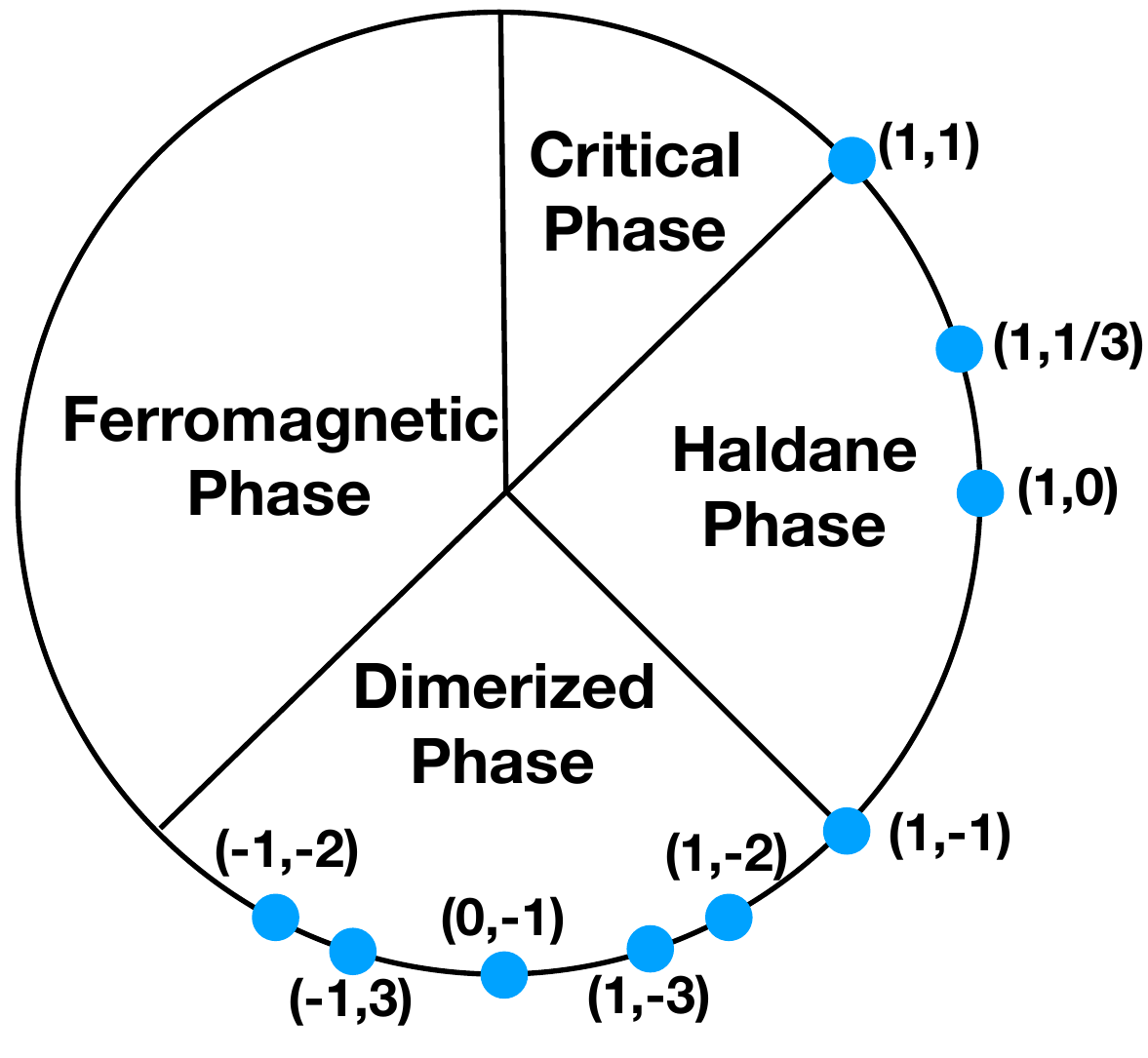}
\caption{Phase diagram of the bilinear-biquadratic $S=1$ model. Figure reproduced from ref.~\onlinecite{liu2012gutzwiller}.}
\label{fig:blbq_phase}
\end{figure}

The physics of the 1D quantum Heisenberg spin-chain is qualitatively different for different spin representations\cite{haldane1983continuum};  half-integer spins have a gapless ground state and power-law spin correlations; integer spins have a gapped ground state with exponentially decaying correlations, the Haldane/AKLT phase\cite{affleck1988valence}.   This latter phase is robust due to a combination of symmetries which protect its topological properties\cite{gu2009tensor,pollmann2012symmetry}.  This symmetry protection can be understood in terms of ``fractionalization'': a $S=1$ spin effectively splits into two $S=1/2$ edge modes that transform under non-trivial projective representations of the symmetries (the product of the symmetry representations differs from the representation of the product). These features are reflected by non-trivial degeneracies in the entanglement spectrum \cite{turner2011topological,pollmann2010entanglement} - i.e. the eigenvalues of $H_\textrm{ent}$ in $\rho_{A} = \text{e}^{-H_\textrm{ent}}$ where  $\rho_{A} = Tr_{B}\ket{\psi}\bra{\psi}$ is the reduced density matrix on an A subsystem \cite{li2008entanglement,levin2006detecting, kitaev2006topological}.  The BLBQ model has four phases as shown in fig.~\ref{fig:blbq_phase}.  This includes the Haldane phase (at the Heisenberg point), as well as a dimerized and critical phase.  

One can derive the relevant projected mean field state from the slave-fermion construction by  fractionalizing the spin operators $\hat{\textbf{S}}$ in terms of fermionic parton operators: 
\begin{equation}
\hat{\textbf{S}}_i = f^{\dagger}_{i;\alpha} \textbf{S}_{\alpha\beta} f_{i;\beta}
\end{equation}
where $f^{\dagger}_{i;\alpha}$ is the $\alpha$-flavor fermionic parton creation operator at site $i$ and $\textbf{S}_{\alpha\beta}$ are the matrix elements of the spin operators in a given representation $S$ \cite{liu2010fermionic}.  Substituting these expressions into the original Hamiltonian gives a quartic fermionic Hamiltonian $H_f$ which can be decoupled through a mean-field.  The resulting mean-field ground state must then be projected back into the original Hilbert space essentially `glueing' together the fractionalized degrees of freedom. 

The slave-fermion construction of the Bilinear-Biquadratic model was studied in ref.~\onlinecite{liu2012gutzwiller} where the authors used VMC to optimize the $(\chi,\delta,\mu)$ parameters of the projected wavefunction and studied the energy of this slave-fermion state compared to the exact energy achieved by TEBD.   In this section, we will take the same points as studied in ref.~\onlinecite{liu2012gutzwiller,liu2014gutzwiller}, convert the slave-fermion states to MPS, and compute the entanglement spectra and the energies.  

At the level of the Gutzwiller projected groundstates, the two gapped phases, the dimer phase and the AKLT phase, are distinguished by their ``fingerprint'' in the low-lying structure of the  entanglement spectrum: the lowest level of the entanglement spectrum of the dimerized topologically trivial phase is singly degenerate (for between dimer cuts); the lowest level of the entanglement spectrum of the Haldane phase ground state is doubly degenerate, corresponding to the presence of two boundary $S=1/2$ edge modes.

\subsection{Generating the MPS}

The relevant mean-field Hamiltonian that arises from the parton construction of the bilinear-biquadratic S=1 model \cite{liu2010fermionic},\cite{liu2014gutzwiller} is
\begin{multline}
H_{mf} = -J\chi\sum_{i,\alpha = -1,0,1} \left[ c^{\dagger}_{i,\alpha}c_{i,\alpha} + \text{h.c.} \right] \\
+(J-K)\Delta \sum_{i,j}\left[c^{\dagger}_{i,-1}c^{\dagger}_{j,1} - c^{\dagger}_{0i}c^{\dagger}_{0j} + c^{\dagger}_{1i}c^{\dagger}_{-1j} +\text{h.c.} \right] \\
+\lambda \sum_{i,\alpha}c^{\dagger}_{i\alpha}c_{i\alpha}
\end{multline}
where the $c^{\dagger}_{-1}$, $c^{\dagger}_{0}$ and $c^{\dagger}_{1}$ are the on-site fermion parton flavors corresponding to $S_z=-1,0,1$.

This could be converted into a MPS by treating it as a general Pfaffian and then applying the techniques in section ~\ref{subsec:pffmps}. In models such as this, though, where the mean-field Hamiltonian in the parton basis has a tensor sum structure where one or more of the Hilbert subspaces can be treated with a simpler mean-field (i.e. with a SD or BdG ground state)  it makes computational sense to obtain the MPS representation in each sector and then ``glue'' the two MPS together; we exemplify this approach here.

Introducing a Nambu spinor, in the $k$-basis the Hamiltonian is block-diagonal:

\begin{gather}
 H^{k}_{mf}
 =
 \frac{1}{2}
  \begin{bmatrix}
  c^{\dagger}_{k,1} & c_{-k,-1} & c^{\dagger}_{k,0} &c_{-k,0}
  \end{bmatrix} \nonumber
  \times \\
  \begin{bmatrix}
   \chi_k & \Delta_k & 0 & 0\\
   \Delta_k^{*} & -\chi_k & 0 & 0\\
   0 & 0 & \chi_k & -\Delta_k  \\
   0 & 0 & -\Delta_k^{*} & -\chi_{k} 
   \end{bmatrix}
   \begin{bmatrix}
   c_{k,1} \\ 
   c^{\dagger}_{-k,-1}\\
   c_{k,0} \\
   c^{\dagger}_{-k,0} 
   \end{bmatrix}
\end{gather}

\noindent

The one-body hamiltonian is a tensor sum of BdG-like Hamiltonian $H_{BdG}$ and a $p+ip$ Hamiltonian $H_{p+ip}$. 
The mean field ground state is (we consider anti-periodic boundary conditions and an even number of sites):
\begin{multline}
\ket{\Psi_{GS}} = \Pi_{0<k<2\pi} \left(u_k+v_k c^{\dagger}_{k,1}c^{\dagger}_{-k,-1} \right) \times \\
\Pi_{0<q<\pi} \left( u_q-v_q c^{\dagger}_{q,0}c^{\dagger}_{-q,0} \right) \ket{0}
\end{multline}
where $u_k$ and $v_k$ are given in terms of the parameters of the Hamiltonian.

By performing a particle-hole transformation in the $S_z = \{\uparrow, \downarrow\}$ sector (see section \ref{subsec:bdgmps}) and the Pfaffian artificial extension in the $S_z = 0$ sector (see section \ref{subsec:pffmps}),
\begin{align}
    f^{\dagger}_{1,k} &= c^{\dagger}_{k,1} \\
    f^{\dagger}_{2,k} &= c_{-k,-1} \nonumber \\ 
    f^{\dagger}_{3,k} &= c^{\dagger}_{k,0} \nonumber \\ 
    f^{\dagger}_{4,k} &= c_{-k,0} \nonumber
\end{align}
where  $\{ f_{k,\alpha}, f_{q,\beta}\} = \delta_{\alpha\beta}\delta_{kq}$,
giving us 
\begin{multline}
    \ket{\Psi_{GS}^{ext}} = 
    \Pi_k \left(u_k f^{\dagger}_{k,2}+v_k f^{\dagger}_{k,1} \right) \\
    \Pi_{0<q<\Pi} \left( u_qf^{\dagger}_{q,4} - v_q f^{\dagger}_{k,3}\right) \ket{\text{vac}}
\end{multline}
with $\ket{\text{vac}} =\Pi_{0<k<2\pi}c_{k\downarrow}\ket{0} \Pi_{0<q<\pi} c_{k,0}\ket{0}$. Since $\ket{\Psi^{ext}_{GS}}$ is a Slater Determinant,  we can obtain the MPS representation using the methods in section \ref{sec:SDMPS}. We now ``undo'' the transformation (see again section \ref{subsec:bdgmps} and section \ref{subsec:pffmps})
and write the MPS in the following form:
\begin{multline}
\ket{MPS} = \sum_{\{i\}} \left( C^{[1_{\uparrow}] i_{1\uparrow}}C^{[1_{\downarrow}] i_{1\downarrow}}
C^{[1,\rightarrow]i_{1,\rightarrow}} \right) \times \\
\ldots
\left(C^{[N_{\uparrow}] i_{N\uparrow}}C^{[N_{\downarrow}] i_{N\downarrow}}C^{[N,\rightarrow]i_{N_\rightarrow}}\right) \\
\ket{\left(i_{1_\uparrow}i_{1_\downarrow}i_{1_\rightarrow}\right)\ldots
\left(i_{N_\uparrow}i_{N_\downarrow}i_{N_\rightarrow}\right)}
\end{multline}
where $i_{n,\alpha} \in \{0,1\}$ indicates the absence/presence of a particle of type $\alpha$ on site $n$.

To obtain the MPS with onsite tensors $A^{[n]\sigma_n}, \sigma_{n} \in \{\uparrow, \downarrow, \rightarrow,  \uparrow\rightarrow, \downarrow\rightarrow, \uparrow\downarrow, \uparrow\downarrow\rightarrow\}$,  we "glue" together appropriate sectors. For example
\begin{equation}
A^{[n]\uparrow} = C^{[n\uparrow]1}C^{[n\downarrow]0}C^{[n\rightarrow]0}
\end{equation}
Gutzwiller projection is realized by summing \textit{only} over the one-particle per site physical indices $\sigma_n \in \{\uparrow, \downarrow, \rightarrow\}$:
\begin{equation}
    P_{G}\ket{\Psi_{GS}} = \sum_{\sigma_n \in \{\uparrow, \downarrow, \rightarrow\}} 
    A^{[1]\sigma_1}\ldots
    A^{[N]\sigma_N} \ket{\sigma_1  \ldots \sigma_N}
\end{equation}
\subsection{iMPS orthogonalization}\label{sub:imps_ortho}

In this section, we will discuss orthogonalizing our iMPS states.  This includes a brief overview of the standard iMPS orthogonalization as well as a detailed description of how we address the degeneracies that appear when Gutzwiller projecting slave-fermion mean-field states onto degenerate ground state manifolds.  

The orthogonalization procedure for a typical iMPS is standard (see ref.~\onlinecite{orus2008infinite} and supplement ~\ref{supplement:ortho_procedures} for more details). The method relies on obtaining the leading right/left eigenvectors of the transfer matrix operator $E = \sum_{\sigma} A^{\sigma} \otimes A^{\sigma*}$ where $\sigma$ runs over the on-site physical index.
$E$ admits the following decomposition:
\begin{equation}\label{eqn:e_decomposition}
E = \sum_i \lambda_i \ket{R}_i \bra{L}_i
\end{equation}
where $\ket{L}_i$ and $\ket{R}_i$ are left/right eigenvectors of $E$ and $\braket{L_i|R_i} = 0$ for $\lambda_i$ non-degenerate. In the infinite limit only the leading left/right eigenvectors of $E$ survive. If the dominant eigenvalue is non-degenerate, the transfer matrix is given by
\begin{equation}\label{eqn:pure_transfer}
    \lim_{N\rightarrow \infty} E^N = \ket{R} \bra{L}
\end{equation}
where $\ket{R(L)}$ are by definition the eigenvectors corresponding to the dominant eigenvalue. Thus, the dominant left and right eigenvectors correspond to a pure state.
The Implicitly Restarted Arnoldi Method can be efficiently used for this purpose by noting that $\left(\sum_{\sigma} A^{\sigma} \otimes A^{\sigma*}\right) \mathrm{vec}(v) = \sum_{\sigma} \mathrm{vec} \left( A^{\sigma*} v A^T \right)$, where the $\mathrm{vec}(v)$ operation takes the square matrix $v$ and stacks the columns together. The entanglement spectrum and observables are then easily obtained.

When the leading eigenvalues of the transfer matrix are degenerate in magnitude 
\begin{equation}\label{eqn:mixed_transfer}
\lim_{N\rightarrow \infty} E^N = \ket{R_1} \bra{L_1} + \ket{R_2} \bra{L_2} 
\end{equation}
$E^N$ is in mixed form. In general the output of the Arnoldi method gives $\braket{L_i|R_i} \neq 0$. 
Thus, additional steps are required to obtain the canonical form of the iMPS. The degeneracy of the leading eigenvalues signals the presence of degenerate states. This is indeed what happens for the 2-fold degenerate dimer phase and the 4-fold degenerate Haldane phase (in the thermodynamic limit). In order to access all the states in the ground state manifold, we need to obtain the proper set of pure iMPS states. The transfer matrix of a pure iMPS has unique left/right leading eigenvectors. 

Here we consider the case of two-fold degeneracy present in the dimer phase states (other states and higher degeneracies can be dealt with using a similar procedure).  For a two-fold  degenerate iMPS, we need to find two pure iMPS generated by bulk tensors $A_1$ and $A_2$.  Any iMPS within the degenerate manifold is then able to be written as a linear superposition of these pure states: $\ket{\psi_{B}} = \alpha_1 \ket{\psi(A_1)} + \alpha_2 \ket{\psi(A_2)}$ where the notation $\ket{\psi(B)}$ indicates the iMPS generated by bulk tensor $B$.  Note that the entanglement spectrum of the reduced density matrix $\rho_B = |\alpha_1|^2 \rho_{A_1} + |\alpha_2|^2 \rho_{A_2}$ is given by the combined spectra of $\alpha_1 \rho_1$ and $\alpha_2 \rho_2$.

To generate these pure iMPS, we start from a non-canonical bulk tensor $A$  iMPS with a single site unit cell (typically generated by projection).  In the case of the dimer phase, this bulk tensor has two leading right (respectively left) eigenvectors, $v_1$ and $v_2$, with equal magnitude eigenvalues $|\eta_1| = |\eta_2|$, but different signs (i.e. $\eta_1 = -\eta_2$).  

According to Theorem 5 in ref.~\onlinecite{perez2006matrix} and Theorem 11 in ref.~\onlinecite{ruiz2011tensor}, there is a unitary that transforms each of the matrices
$B^{\sigma\sigma'} = A^{\sigma}A^{\sigma'}$ into block diagonal form, with two blocks; the two blocks are the 2-site uniform tensors corresponding to the two pure states.

\begin{figure}[!h]
    \includegraphics[width=0.5\textwidth]{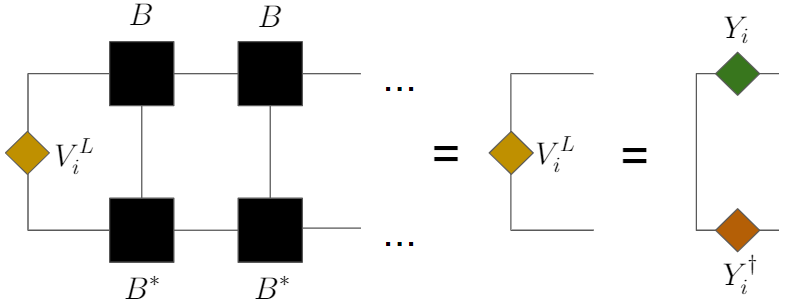}
    \includegraphics[width=0.5\textwidth]{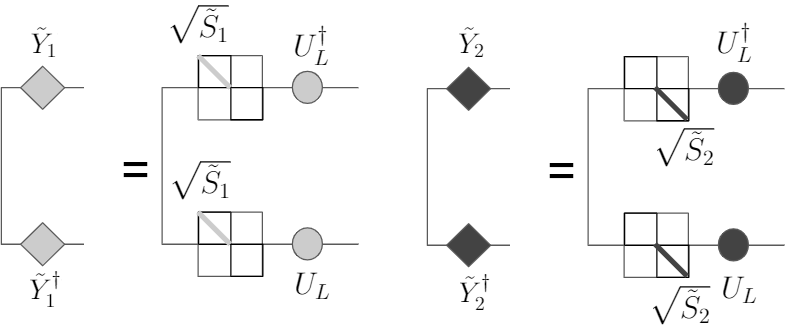}
    \includegraphics[width=0.5\textwidth]{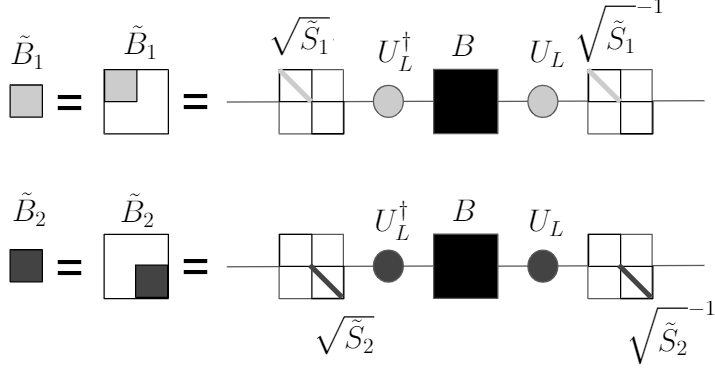}
    \includegraphics[width=0.5\textwidth]{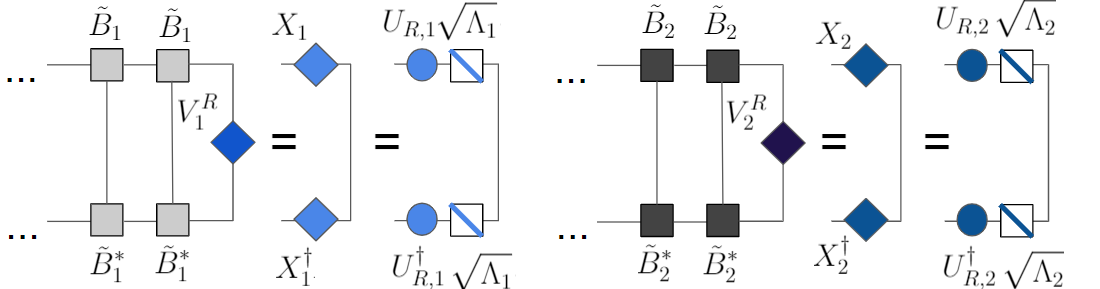}
    \includegraphics[width=0.5\textwidth]{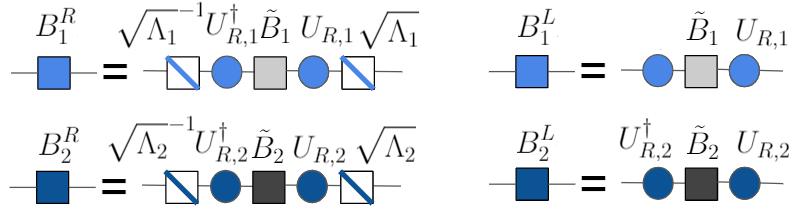}
    \caption{Illustration of decomposition of a mixed transfer matrix into pure components.
    In step 1 (first line) we find the dominant left-eigenvector of the transfer matrix;
    In step 4 (second line), we find $U_L$ and $\tilde{S}_1$ and $\tilde{S}_2$.
    In step 5 (third, fourth line and fifth line), we form two new tensors, $\tilde{B}_i$ and find their right leading eigenvectors.
    In step 6 (other lines), we find the entanglement spectrum $\sqrt{\Lambda_i}$ and the right/left canonical matrices, $B^R_i$ and $B^L_i$, corresponding to the two pure states. 
    \label{fig:iMPSDegeneracy} }
\end{figure}

Based on the mathematical  theorems in refs.~\onlinecite{perez2006matrix} and  \onlinecite{ruiz2011tensor}, we use the following procedure to compute the pure states (see fig.~\ref{fig:iMPSDegeneracy}): 
\begin{enumerate}
\item Start with the $D \times D$ ($D$ is the bond dimension of the bulk tensors $A$) left leading eigenvectors (with the same eigenvalue), $V^L_1$ and $V^L_2$, of the completely positive map $E^2$, i.e. $\sum_{\sigma,\sigma'}B^{\dagger\sigma\sigma'}V^L_i B^{\sigma'} = V^L_i$.

\item $V^L_i$ are transformed into hermitian matrices: $V^L_i := 1/2 (V^L_i + (V^L_i)^{\dagger})$; this is possible because if $V^L_i$ is an eigenvector of $E^2$, then $(V^L_i)^{\dagger}$ is also an eigenvector and so their sum is Hermitian. If $V^L_i = U_i D_i U^{\dagger}_i$, then we can write $V^L_i = Y^{\dagger}_iY_i$ with $Y_i = \sqrt{D_i}U^{\dagger}_i$.

\item Diagonalize $V^L_1$ and $V^L_2$ together; this can be done since $[V^L_1, V^L_2] = 0$ so that $U^{\dagger}V^L_i U = S_i$, with $S_i$  being a diagonal matrix.

\item Form two linear combinations $\tilde{V^L_i} = V^L_1 - \alpha_i V^L_2$ where $\alpha_i$ is one of the two non-zero values obtained by the elementwise division of $S_1$ and $S_2$.
Then $U^{\dagger}_L \tilde{V^L_1} U_L $ will be a diagonal matrix $\tilde{S_1}$ with entries
$\left(\tilde{d}^1_1,\tilde{d}^2_1,\ldots,\tilde{d}^p_1,0,0,\ldots 0\right)$ and $U^{\dagger}_L\tilde{V^L_2} U_L $ a diagonal matrix $\tilde{S_2}$ with entries $\left(0,0,\ldots,0,\tilde{d}^{D-k}_2,\tilde{d}^{D-k-1}_2,\ldots,\tilde{d}^{D}_2,\right)$ and $D-k \geq p$; in fact, it will almost always be the case that $D-k>p$, since the bond dimension of the canonical bulk tensor decreases after projection; this decomposition is guaranteed by Theorem 5 in ref.~\onlinecite{perez2006matrix}. 
\item Form two new $2$-site bulk tensors $\tilde{B}_i = \sqrt{\tilde{S}_i}U^{\dagger}_L B U_L \sqrt{\tilde{S}_i}^{-1}$ and obtain their transfer matrix right leading eigenvectors; they will each have a unique leading hermitian semi-positive definite diagonal eigenvector, $V^R_i = U_{R,i} \Lambda_i \left(U_{R,i}\right)^{\dagger}$; we can write $V^R_i = X_i X^{\dagger}_i$ with $X_i = U_{R,i} \sqrt{\Lambda}$; then $\sqrt{\Lambda_i}$ is the entanglement spectrum of the corresponding pure state.

\item  $B^R_i = \sqrt{\Lambda_i}^{-1} \left(U_{R,i}\right)^{\dagger} \tilde{B}_i U_{R,i} \sqrt{\Lambda_i} $ are the right canonical tensors and $B^L_i = \left( U_{R,i} \right)^{\dagger}\tilde{B}_i U_{R,i}$ are the left canonical tensors; since $B^L_i \sqrt{\Lambda_i} = \sqrt{\Lambda_i}B^R_i$ the uniform $2$-site translationally invariant bulk tensors can be written as $A_i = \sqrt{\sqrt{\Lambda_i}}B^R_i \sqrt{\sqrt{\Lambda_i}}^{-1} = \sqrt{\sqrt{\Lambda_i}}^{-1} B^L_i\sqrt{\sqrt{\Lambda_i}}$.
\end{enumerate}

\begin{table*}[t]
  \centering
  \begin{tabular}{ |c|c|c|c|c|c|c|c|c|c|c|}
    \hline
 & (1,1)$_{ULS}$ & (1,$\frac{1}{3}$)$_{AKLT}$ & (1,0)$_{Heisenberg}$ & (1,-1)$_{TB}$ &  (1,-2) & (1,-3) &(0,-1) & (-1,-3) & (-1,-2)  \\
\hline
itebd [\onlinecite{liu2012gutzwiller}] &0.2971 & $-\frac{2}{3}$ &-1.4015 &-4 & -6.7531 & -9.5330 &-2.7969 &-7.3518 &-4.5939  \\
\hline
dmrg & 0.2978  & $-\frac{2}{3}$  &-1.4015 & -3.9999 & -6.7526 &-9.5314  &-2.7969  & -7.3516 & -4.5939 \\
\hline
VMC  [\onlinecite{liu2012gutzwiller}] & 0.2997 & $-\frac{2}{3}$  & -1.4001 & -3.9917  & -6.7372 & -9.5103  & -2.7953 & -7.2901 & -4.4946 \\ 
& $\pm 0.0004$ & $\pm 7 \times 10^{-15}$ & $\pm 0.0004$ & $\pm 0.0012$ & $\pm 0.0023$ & $\pm 0.0034$ & $\pm 0.0005$ & $\pm 0.0038$ & $\pm 0.0028$ \\ 
\hline
fMPS & 0.2995 & $-\frac{2}{3}$ &-1.3999 &-3.9895 & -6.7369 &-9.5073 &-2.7948 & -7.2877 &-4.4935 \\
\hline
iMPS &---  &$-\frac{2}{3}$ & -1.3999 &---& -6.7368 &-9.5071 & -2.7947  & -7.2877 & -4.4934 \\
\hline
$\chi$ & 1  & 1 & 1& 1  & 1  & 1 & 0& 0 & 0   \\
\hline
$\Delta$ & 0  & $\frac{3}{2}$ & 0.98 &  1.11  & 1.15  & 1.79 & 1 &  1& 1  \\
\hline
$\lambda$ & 1 & 0 & 1.78 &  2.00  & 2.07 & 2.22 & 0.14 &  0.21 & 0.12  \\
\hline
  \end{tabular}
  \caption{Comparison of energies per site between the exact ground state (iTEBD/DMRG), variational Monte Carlo (VMC) and the fMPS and iMPS generated from the projected slave-fermion states.  fMPS are computed with $N=64$ or $N=96$ and $10^{-4}\leq \epsilon \leq 3\times10^{-4}$. The bulk tensors used in iMPS have bond dimension $D \approx 1000$.  Column headings correspond to $(J,K)$
 }
 \label{table:energy}
\end{table*}

\subsection{Energy of BLBQ Slave-Fermion Wavefunctions}

We compute both the MPS and iMPS (except at the critical points) for the variational Gutzwiller projected wavefunctions corresponding (as found in ref.~\onlinecite{liu2012gutzwiller} by minimizing the variational energy) to the points in fig. \ref{fig:blbq_phase}.  We directly compare the energy for all of these points (see table \ref{table:energy}) and find that the energies are all within the error bars reported for the VMC calculation\cite{liu2012gutzwiller}.

\subsection{Entanglement Spectra of BLBQ Slave-Fermion Wavefunctions}
\subsubsection{Dimer phase}
In this section, we will consider entanglement spectra of the dimerized phase of the BLBQ model.
The ground state of the dimerized phase is two-fold degenerate depending on whether the dimer covering spans even or odd bonds;  the entanglement spectra also depends on whether the entanglement cut is made through or between dimers.   For the fMPS, we can obtain both the even and odd cut entanglement spectra of the dimerized states by choosing two consecutive cuts whereas for the iMPS we use the procedure described in sec.~\ref{sub:imps_ortho} to find the two pure states which corresponds respectively to the even and odd cuts. 

\begin{figure}[!h]
    \includegraphics[width=\columnwidth]{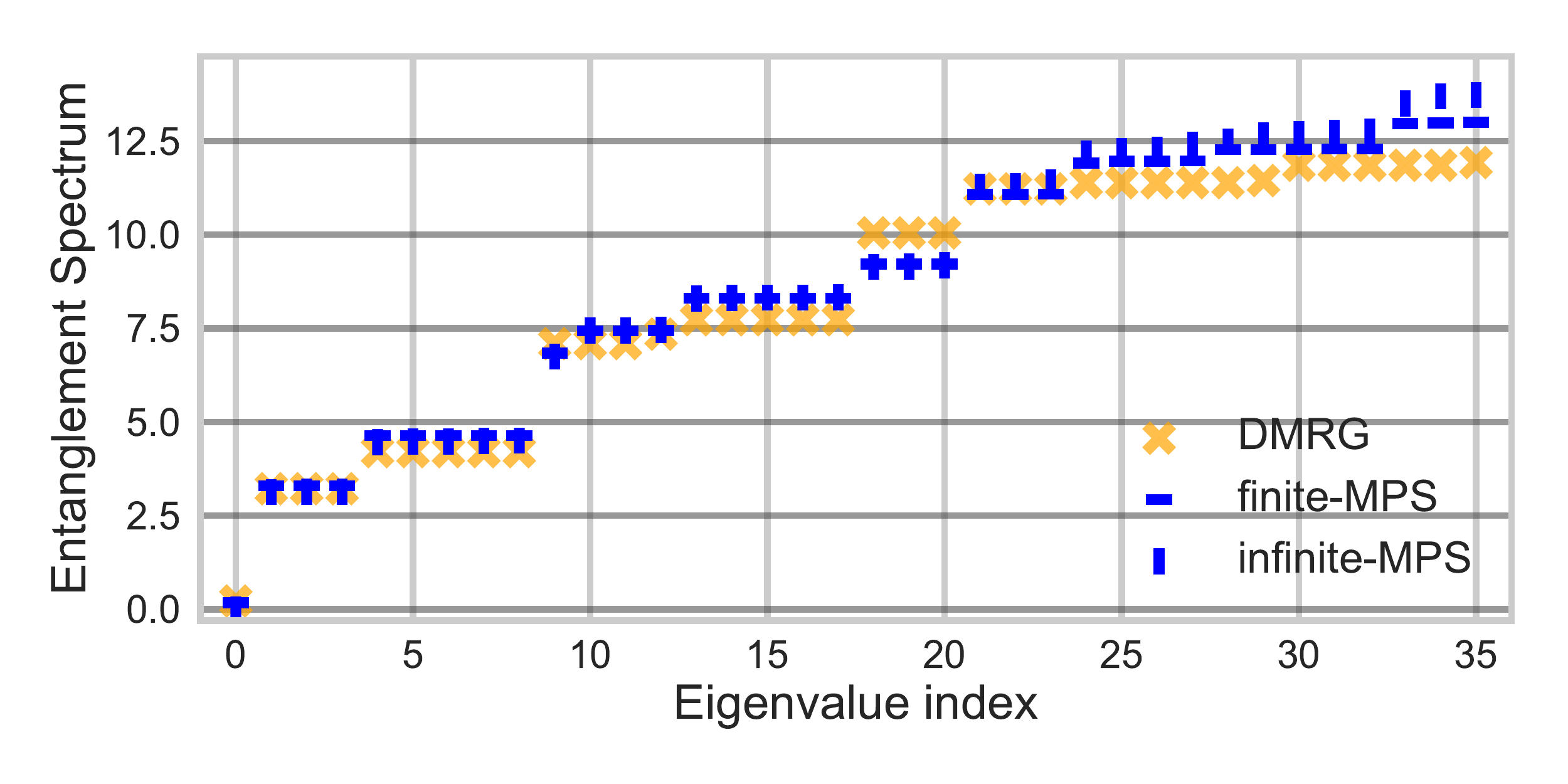}
        \includegraphics[width=0.48\columnwidth]{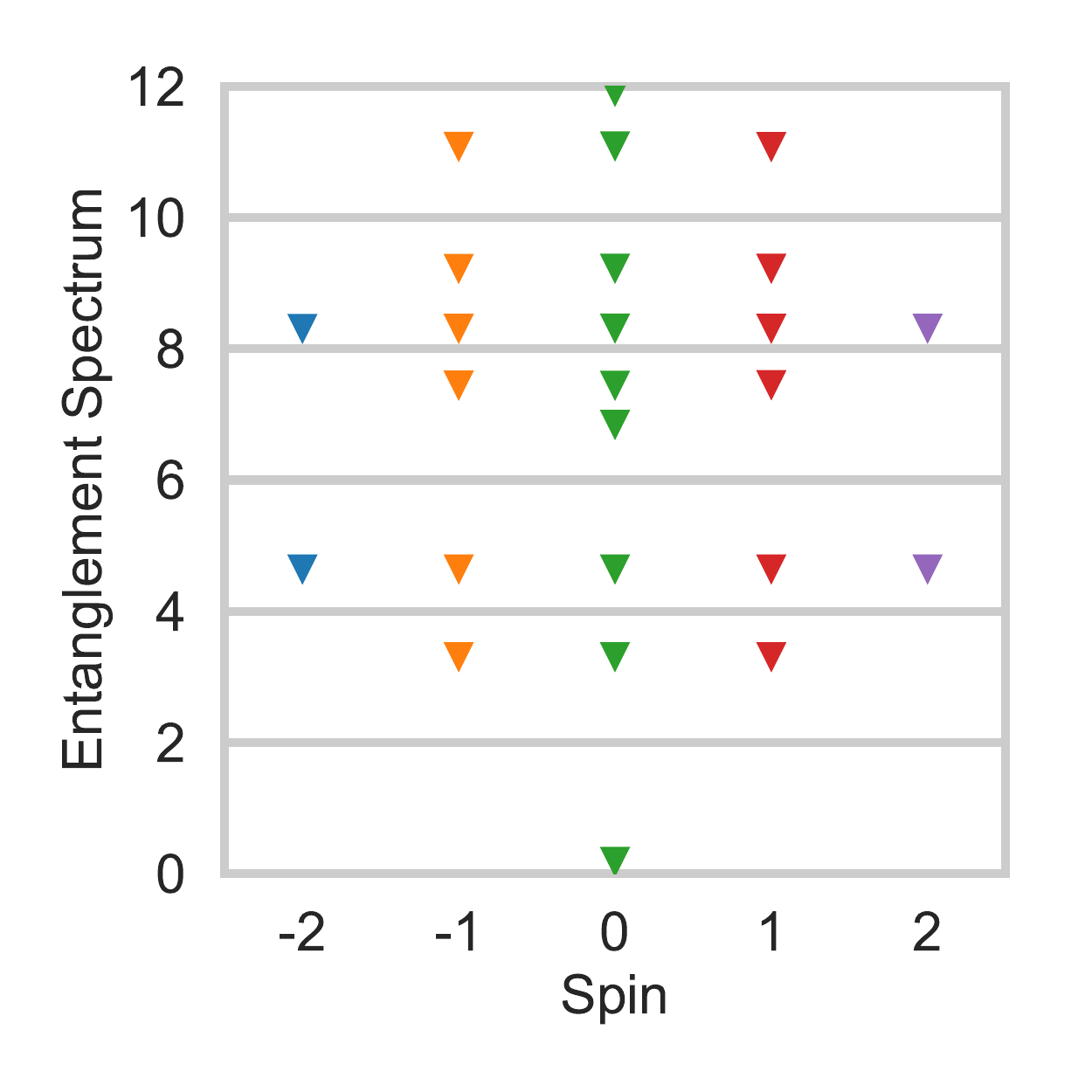}
    \includegraphics[width=0.48\columnwidth]{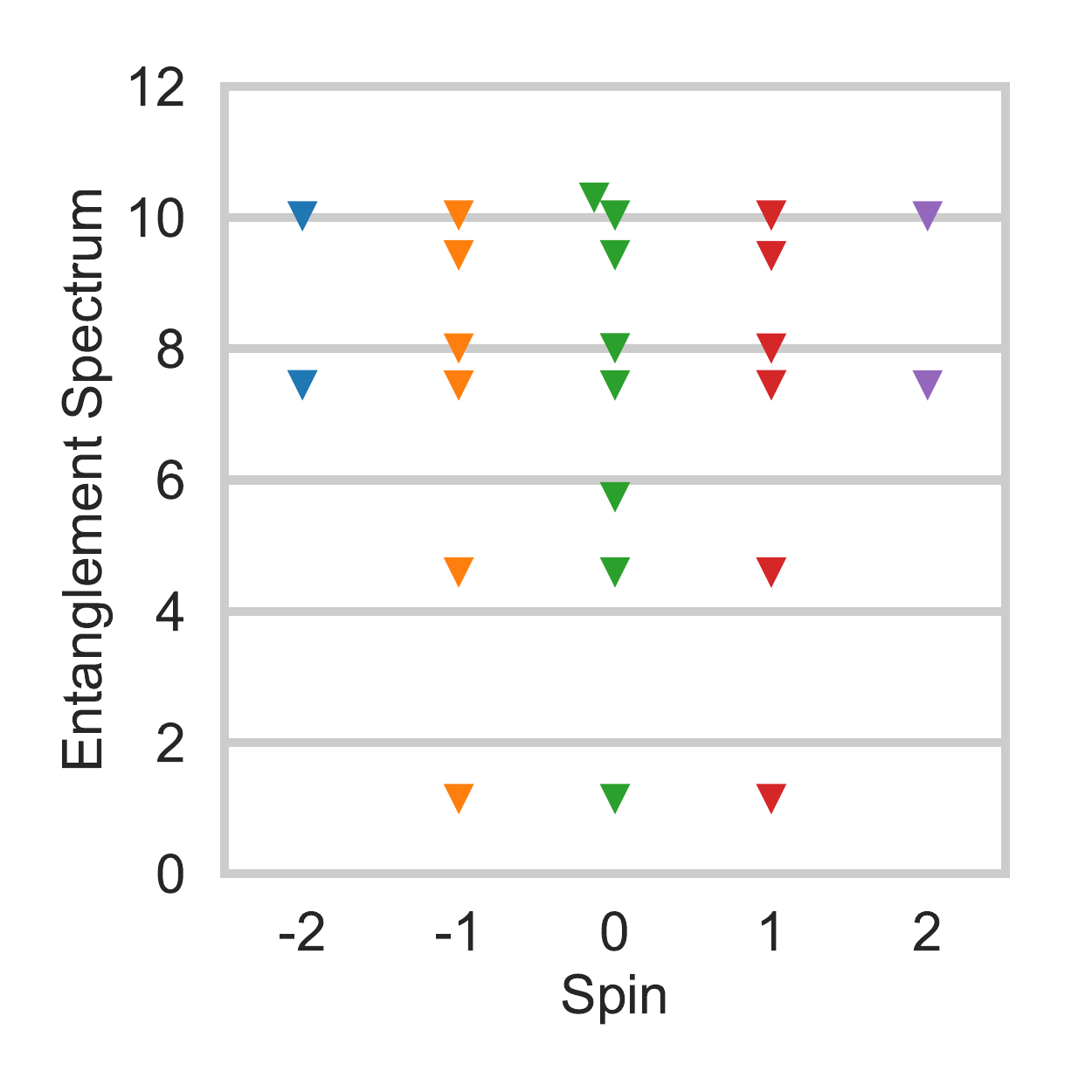}
    \caption{Entanglement spectra of the projected mean-field state from iMPS and fMPS at $(J,K) = (1,-2)$ showing (a) the comparison with DMRG between dimers; and the fMPS spin-resolved ES (b) between dimers and (c) within dimers.  Note that the single and triple degeneracy seen in (b) and (c) are the expected low-level structure seen in a pure VBS state.  The lowest three entanglement levels of (b) are representations of SU(2): singlet, triplet and quintet.}
\label{fig:1m2}
\end{figure}

We start by considering a generic slave-fermion point in the BLBQ model; see fig.~\ref{fig:1m2} for the entanglement spectrum.  The iMPS and fMPS slave-fermion point agrees well both with each other and the exact ES from DMRG.

The low-lying level of the entanglement spectrum cycles between a singlet and a triplet as we move the location of the entanglement cut within the chain. This is indicative of translation invariance breaking and the dimerized structure of the ground state: namely the low-level singlet is associated with a cut between dimers, whereas the low-level triplet is associated with a cut inside dimers. 

We can also further understand the higher states in the entanglement spectra.  A generic point in the dimer phase of the BLBQ model is $SU(2)$ symmetric. Consequently, the entanglement levels transform under $SU(2)$ representation and therefore we expect that degeneracies  should go as the dimension of $SU(2)$ representations (i.e. $2n+1$ for non-negative integer $n$); this can be seen in the multiplet structure of both the entanglement spectra in fig.~\ref{fig:1m2}(bottom-left). From \ref{fig:1m2}(bottom-left) where the lowest three degeneracies between dimers form the singlet($S=0$), triplet($S=1$) and quintuplet($S=2$).

\begin{figure}[!h]
    \includegraphics[width=\columnwidth]{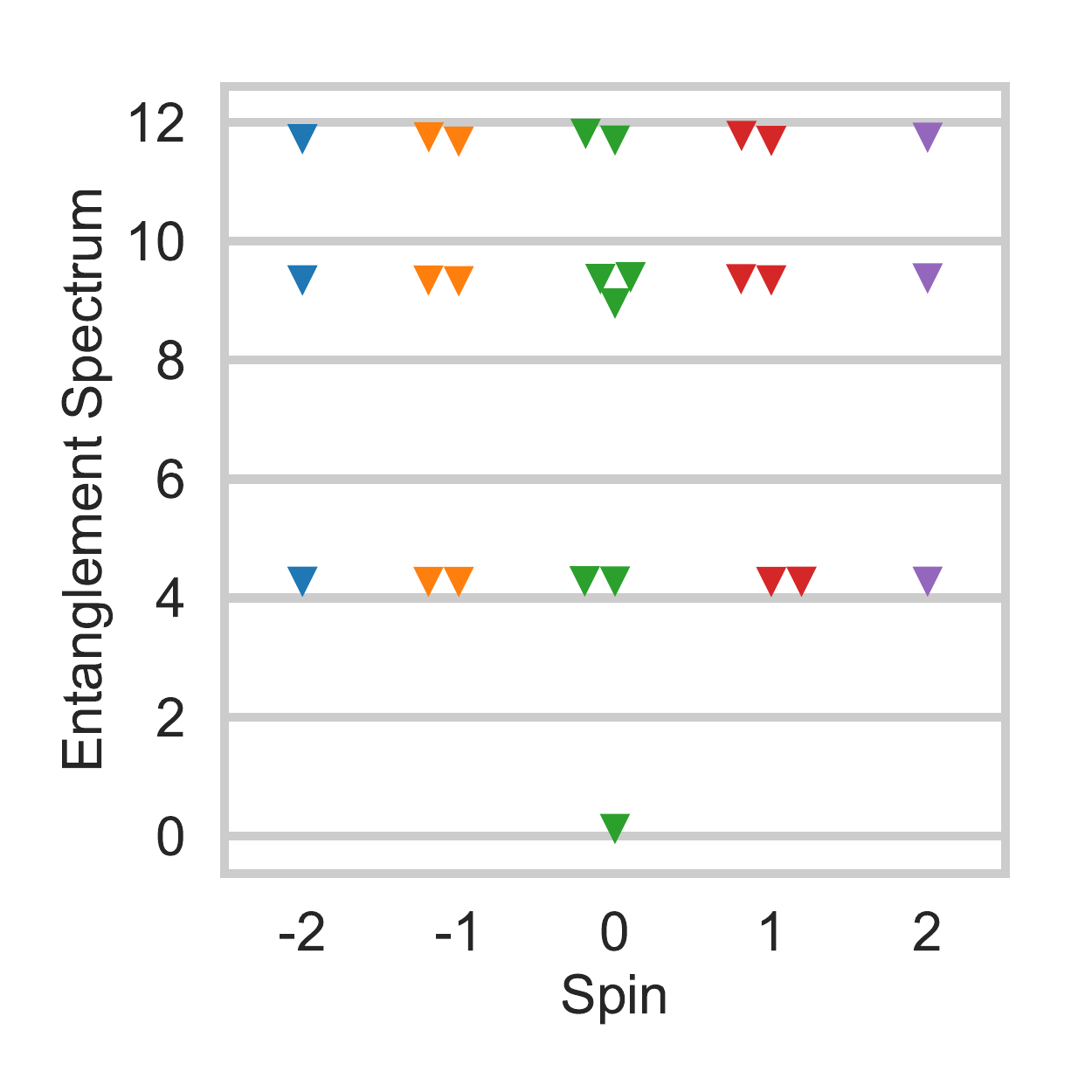}
    \caption{Spin resolved entanglement spectrum (between dimers) for the $(J,K)=(0,-1)$ variational point generated from fMPS. The $SU(3)$ symmetry forces the $S=1$ and $S=2$ into a degenerate octet.}
\label{fig:kbb}
\end{figure}

Beyond considering a generic point within the dimer phase, we now consider the exactly solvable point where $(J,K) = (0,-1)$ (the so-called KBB point \cite{klumper1989new}, \cite{barber1989spectrum}), which is invariant under a larger symmetry group, $SU(3)$ (as opposed to $SU(2)$). This larger symmetry group forces the triplet and quintet to form an octet (the adjoint representation of $SU(3)$). Variationally, the vanishing of the hopping parameter in the slave-fermion mean-field hamiltonian forces this larger symmetry group at the level of the variational Gutzwiller projected wavefunction.  See fig.~\ref{fig:kbb}.  

The points $(J,K)=(-1,-2)$ and $(J,K) = (-1,-3)$ which are found at a variational minima with $t=0$ in the parent Hamiltonian by ref.~\onlinecite{liu2012gutzwiller} also have  $SU(3)$ symmetry.  This symmetry is not present in the true DMRG ground state which transforms only under $SU(2)$ symmetry;  therefore, a better agreement is obtained by perturbing slightly away from this point (see ~\ref{supplement:symm_dimer}).

\subsubsection{Haldane phase}
In this section we compute the entanglement spectra of the AKLT and Heisenberg points belonging to the Haldane phase of BLBQ.  The slave-fermion mean-field for this model has a 4-fold degeneracy at the fermi level that allows for choosing six orthogonal pre-projected mean-field states (at half filling). Projecting each of these states, generates (post-projection) a space of MPS which span 4 degenerate ground states which correspond to the representations of the sum of the two fractionalized $S=1/2$ edge modes of the Haldane phase.  

Here we start by considering the AKLT point which has an exact analytic solution. The AKLT point $(J,K) = (1,1/3)$ is exactly mapped under the above slave fermion projective construction to $(\chi,\delta,\mu) = (1,3/2, 0)$.   To find the AKLT state which (for example) has the edge modes $\uparrow\downarrow$ we can either search in the four-fold projected degenerate space or choose the correct orbitals at the fermi-level pre-projection.  We find the entanglement spectra for each of the four fMPS which correspond to $\downarrow\downarrow$,  $\downarrow\uparrow$, $\uparrow\downarrow$, $\uparrow\uparrow$ edge spin configurations is equal to $\ln(2)$.

For the iMPS, unlike the dimer phase, where there was $2$-fold degeneracy in the leading eigenvalues of the transfer matrix operator for points in the Haldane-phase, we find $4$-fold degeneracy. We obtain $2$-negative and $2$-positive (equal in magnitude) leading eigenvalues.  Taking the space spanned by the two eigenvectors with positive eigenvalues, we apply the iMPS orthogonalization procedure from sec.~\ref{sub:imps_ortho}. From this process, for the AKLT slave-fermion point we find after orthonormalization the iMPS

\begin{equation}\label{eqn:aklt_imps}
\begin{split}
A^{0} &= 
\begin{pmatrix}
-1& 0 \\
0 & 1
\end{pmatrix} \\
A^{\downarrow} &= \sqrt{2}
\begin{pmatrix}
0& -e^{-i\theta} \\
0 & 0
\end{pmatrix} \\
A^{\uparrow} &= \sqrt{2}
\begin{pmatrix}
0& 0  \\
e^{i\theta} & 0
\end{pmatrix}
\end{split}
\end{equation}
associated with two pure states (i.e $\ket{\uparrow\downarrow}$ and $\ket{\downarrow\uparrow}$ of the edge modes). In supplement ~\ref{supplement:mg_chain}, we also obtain the iMPS representation of the $S=1/2$ VBS groundstate of the  Majumdar–Ghosh (MG) chain.

\begin{figure}[!h]
    \includegraphics[width=\columnwidth]{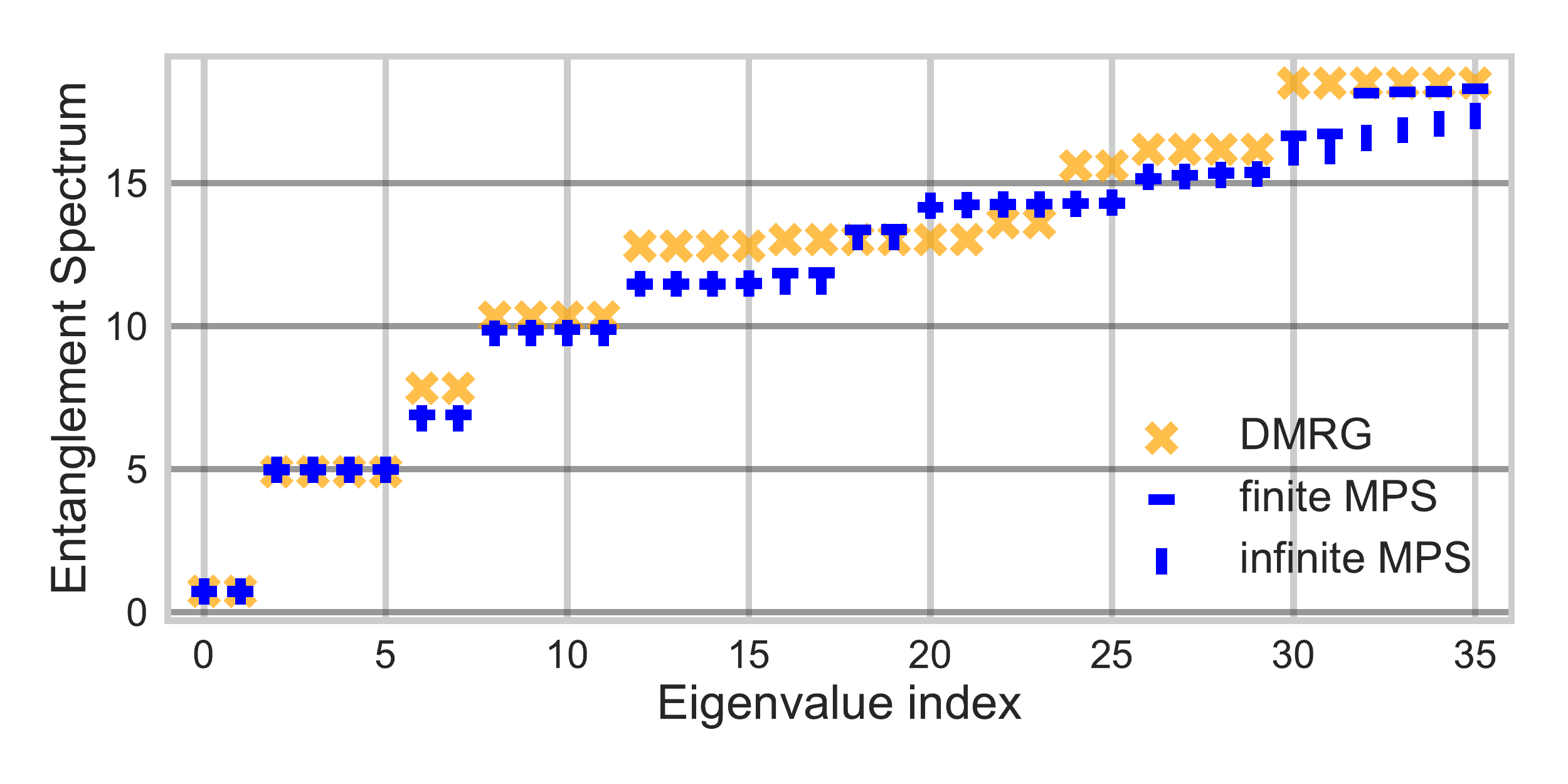}
    \includegraphics[width=\columnwidth]{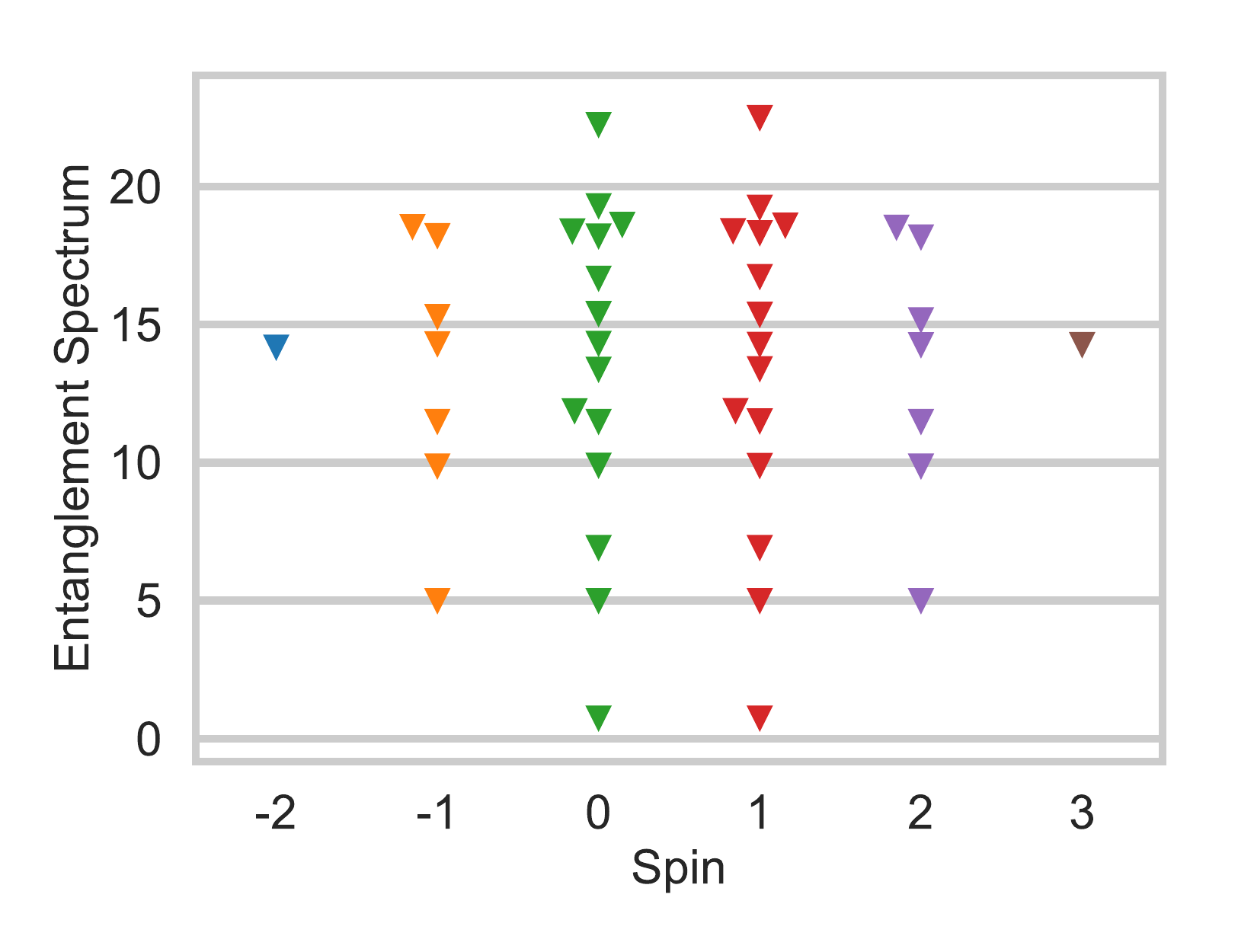}
    \caption{(top) Comparison of entanglement spectrum obtained from fMPS, iMPS and DMRG at the $(J,K) = (1,0)$ Heisenberg point. (bottom) Spin resolved entanglement spectrum from fMPS $(J,K) = (1,0)$ Heisenberg point in the $(S=1,S_z = 1)$ sector.}
\label{fig:heis_es}
\end{figure}

In fig. \ref{fig:heis_es} we also present the entanglement spectrum of the Heisenberg point $(J,K) = (1,0)$ obtained using both fMPS and iMPS. We see that the lower levels match well the entanglement spectrum levels obtained from DMRG of the true Heisenberg ground state. Discrepancies occur naturally higher-up in the spectrum as the variational wavefunction is not the \textit{exact} groundstate. However, the entanglement spectrum of the variational groundstate (qualitatively) captures the symmetries and degeneracies of the \textit{true} entanglement spectrum. Note the fact that every level has even degeneracy comes from the topological nature of the phase.  Moreover, notice that the lowest part of the entanglement spectra match the AKLT state in the same sector. 
\subsubsection{Critical points}
We compute the two critical points at $(J,K)=(1,-1)$ (the Takhtajan-Babujian(TB)  point \cite{takhtajan1982picture},\cite{babujian1982exact}) and at $(J,K)=(1,1)$(the Uimin-Lai-Sutherland (ULS) \cite{uimin1970one},\cite{lai1974lattice}, \cite{sutherland1994model}); they are gapless and hence we analyze them only in the framework of our fMPS method.  The TB ground state is unique; the associated effective conformal field theory is $SU(2)|_{k=2}$.

The ULS ground state is also unique. However, it has an enlarged $SU(3)$ symmetry group; the associated effective conformal field theory is $SU(3)|_{k=1}$. In particular it can be mapped to the $SU(3)$ nearest-neighbor Heisenberg model \cite{thomale2015entanglement}. This enforces an equal number of ``quark'' particle constraints (and hence global equal number of spins $1,-1,0$). At the mean-field level, the pairing parameter vanishes since $J-K=0$. The Hamiltonian is then a tensor sum of 3 identical hopping Hamiltonians acting independently on the fermions of flavour ``up'', ``down'' and ``zero''.  The particle number constraint of $c_1, c_{-1}, c_0$
is naturally enforced at the mean-field level if the the number of sites $N$ is a multiple of $3$. 

\begin{figure}[!h]
    \includegraphics[width=\columnwidth]{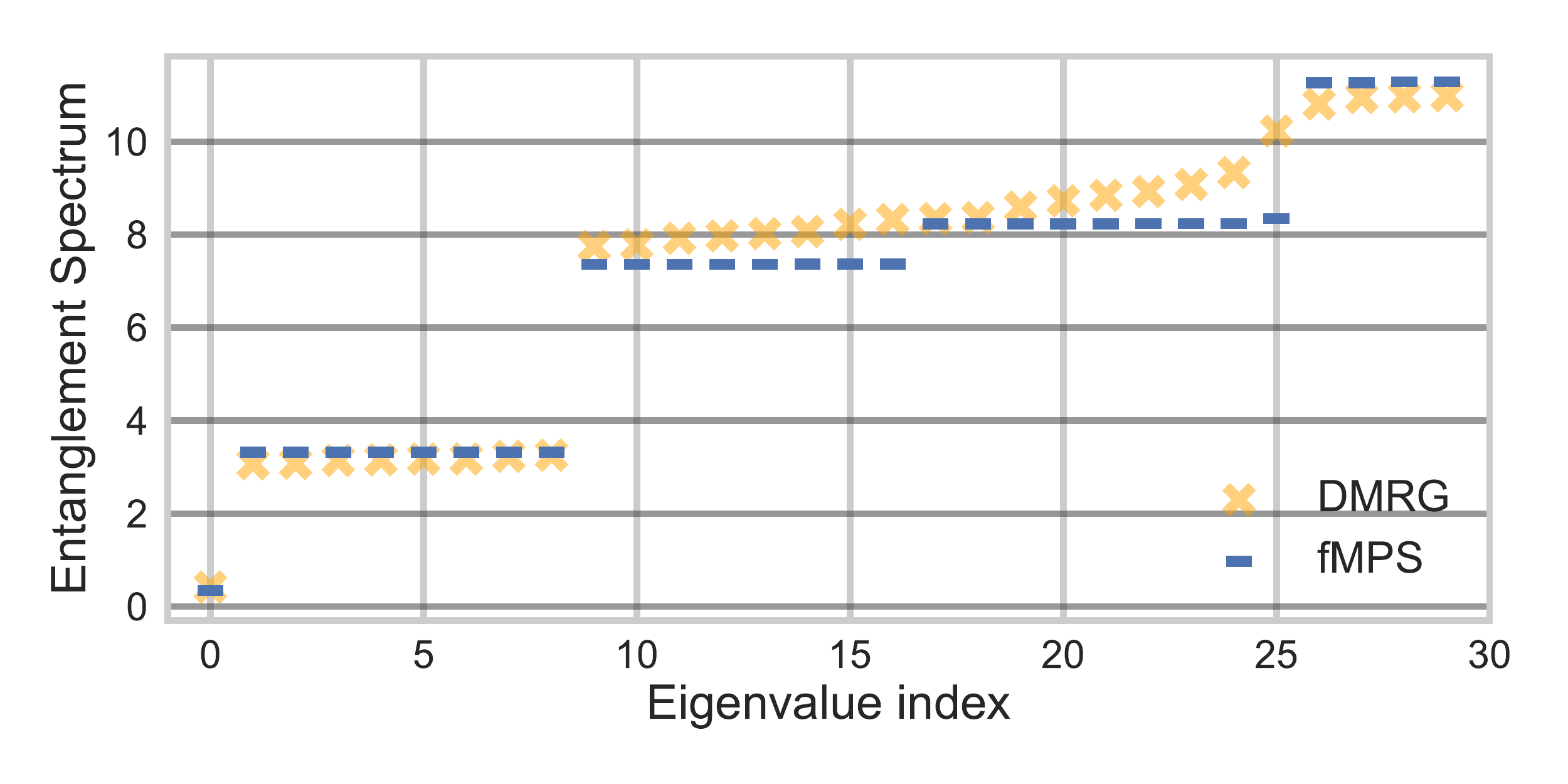}
    \includegraphics[width=\columnwidth]{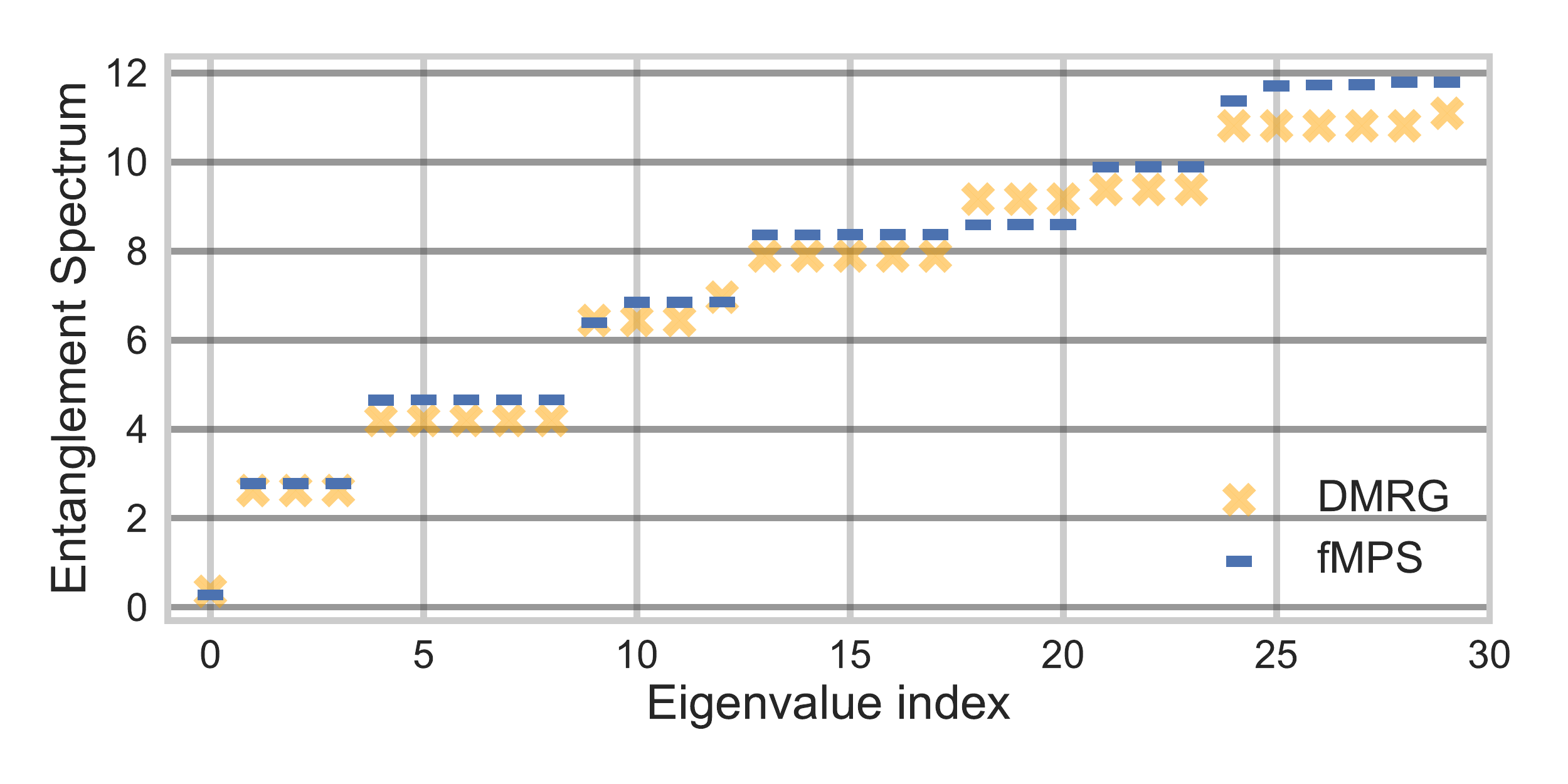}
        \includegraphics[width=0.48\columnwidth]{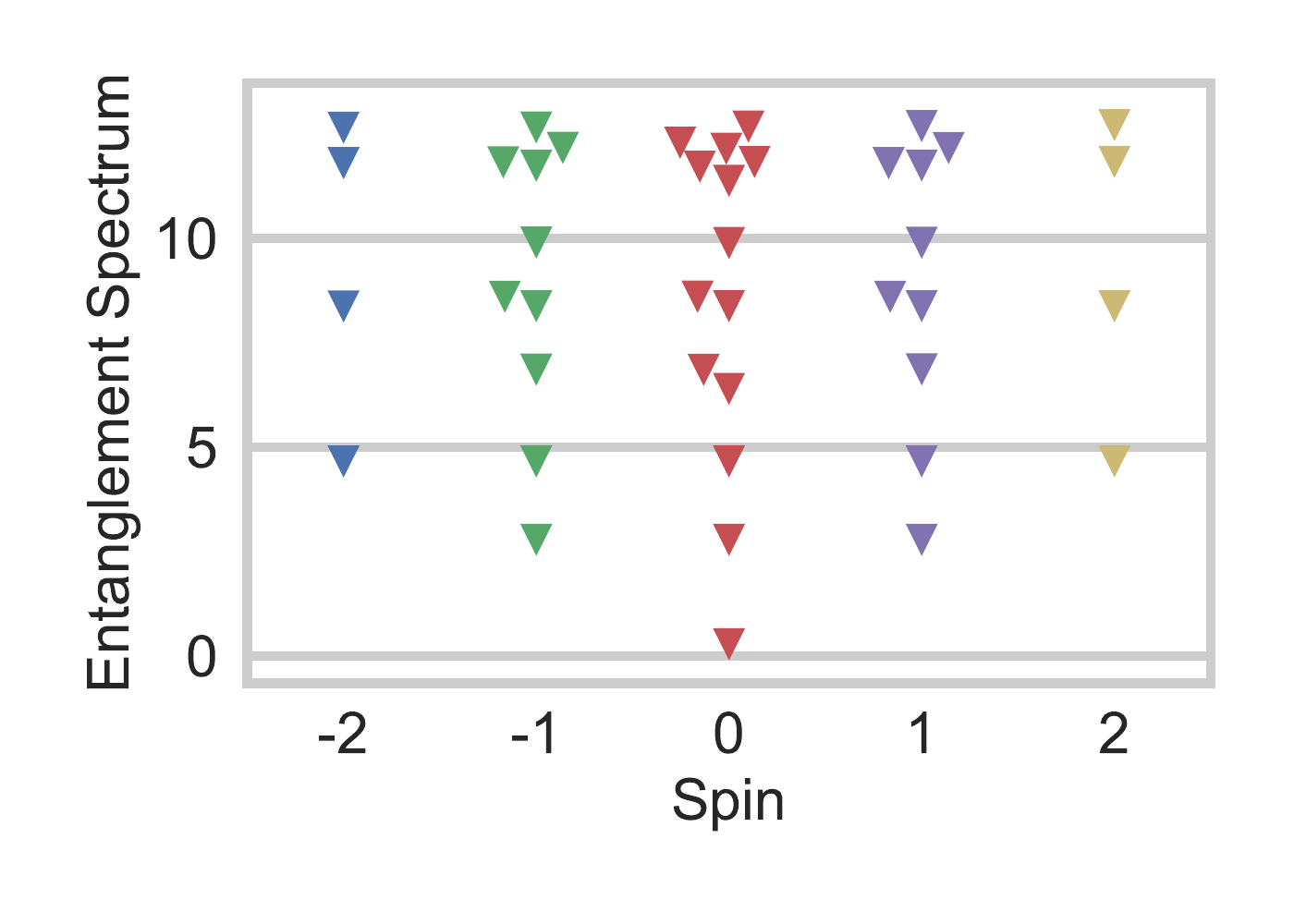}
        \includegraphics[width=0.48\columnwidth]{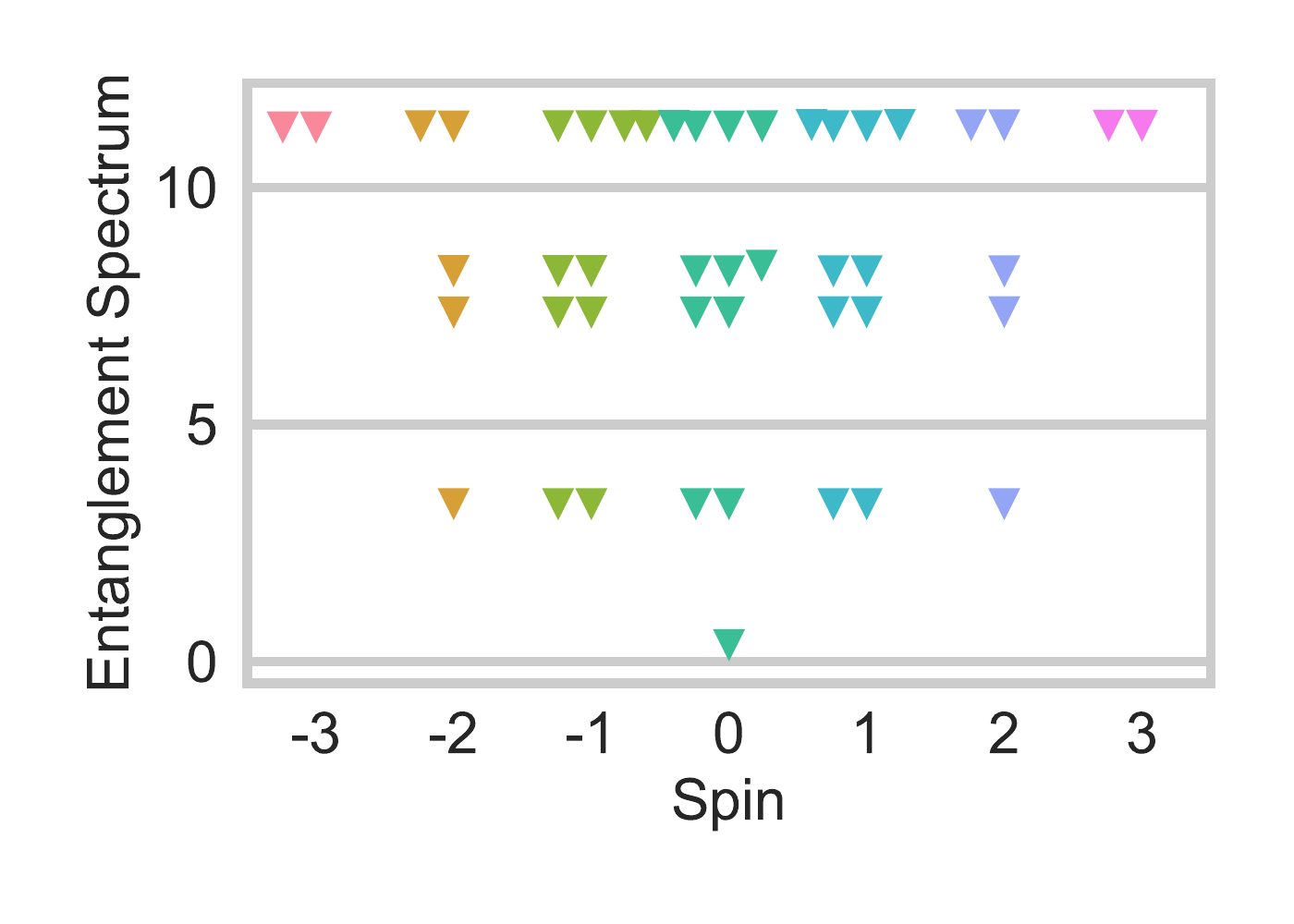}
    \caption{(a) Comparison of entanglement spectra at ULS point for fMPS and DMRG.  (b) Comparison of entanglement spectra at TB point for fMPS and DMRG.  (c) Spin resolved entanglement spectrum for (left) TB point and (right) ULS point}
\label{fig:uls_tb}
\end{figure}

The $SU(3)$ symmetry of the ULS point is reflected in the degeneracies of the entanglement spectrum where the $S=1$ and $S=2$ levels combine together to form $SU(3)$ octets as can be seen in fig.~\ref{fig:uls_tb}. For the TB point the $S=1$ and $S=2$ entanglement levels remain separated.

The central charge of $SU(N)|_k$ CFTs is given by: $c = k \left( N^{2} - 1 \right) /(N + k)$. Hence, analytically $c = 1.5$ for the TB point and $c = 2$ for ULS.
Calabrese and Cardy \cite{calabrese2004j} obtained the following expression for the entanglement entropy scaling for a 1D critical gapless point of finite size $L$ with open-boundary conditions and partition size $x$:
\begin{equation}
S(x,L) = \frac{c}{6} \ln{\frac{2L}{\pi} \sin{(\frac{\pi x}{L})}} + \ln{g} + s_1/2
\end{equation}
where $\ln{g}$ is a boundary entropy term and $S(x,L)$ is the von-Neumann entanglement entropy. 

\begin{figure}[!h]
    \includegraphics[width=\columnwidth]{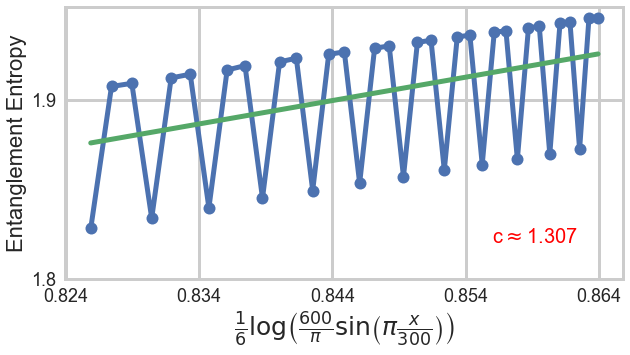}
    \includegraphics[width=\columnwidth]{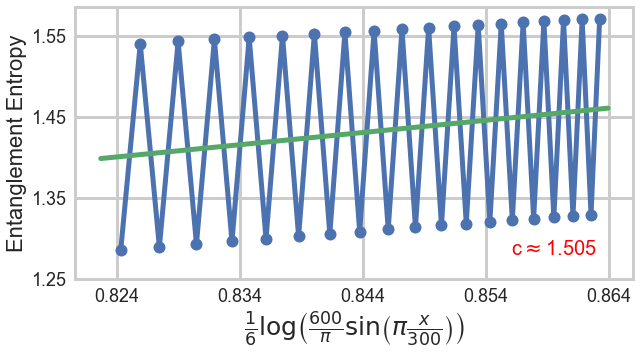}
    
    \caption{Von-Neumann entanglement entropy scaling of the slave-fermion wave-functions describing the ULS (top) and TB (bottom) critical points for a system of size $L=300$ computed with fMPS. Shown also is the central charge obtained from least squares fitting.}
\label{fig:ent_scaling}
\end{figure}

It was found in ref.~\onlinecite{laflorencie2006boundary} that there is an additional alternating term in $S(x,L)$ which decays away from the boundaries. In fig. \ref{fig:ent_scaling} we plot the entanglement entropy against $\frac{c}{6} \ln{\frac{2L}{\pi} \sin{(\frac{\pi x}{L})}}$ for both TB and ULS models. We work on system of size $L=300$ and plot the region $x \in [L/4,3L/8]$ (so $[75,111]$ for TB and $[76,112]$ for ULS). We picked the lower bound at $L/4$ as it is far enough from the boundary and the upper boundary at $3L/8$ to work in the region of the sine curve with $x$ away from $L/2$ where the curve becomes very flat. For the $SU(3)$ ULS point, when $x$ is a multiple of $3$, the highest eigenvalue Schmidt vector contains equal numbers of ``quarks'' and hence is dominant. For cuts at $x = 3k+1,3k+2$ (for integer $k$), the Schmidt vectors cannot satisfy the particle conservation constraint and hence the highest eigenvalue Schmidt vectors are degenerate. A similar situation occurs at the $SU(2)$ TB point  where the $2$-periodicity is easily explained in the dimer picture: for even cuts we cut between dimers, whereas for odd cuts we break dimers and hence split the singlet apart. The alternating term is still significant for the parameters we chose. Hence, it is difficult to reliably extract the central charge. We overlay the lines obtained from least square fitting for both models. For the $TB$ point we obtain the central charge $c \approx 1.505$ which is remarkably close to its true value of $1.5$. For the $ULS$ point we obtain $c=1.307$ which is substantially away from the true value at the $ULS$ point of $c=2$. 
\section{Discussion and Future Work}

We have developed a series of efficient and highly parallel algorithms to obtain the finite and infinite (for gapped states) MPS representation of fermionic mean-field states. Gutzwiller projection is easily implemented by eliminating the doubly-occupied and unoccupied physical sectors of the mean-field slave-fermion MPS tensors. 
We have used these methods to obtain the (i)MPS representation of Gutzwiller projected mean-field states that arise from the variational slave-fermion approach to the $S=1$ Bilinear-Biquadratic (BLBQ) quantum spin chain introduced in ref.~\onlinecite{liu2012gutzwiller}. We first verify that the energies we obtain via both finite MPS and infinite MPS (not applicable to the critical points) for the points considered are within the error bars of their VMC calculations~\cite{liu2012gutzwiller}.

Additionally, we obtain the entanglement spectra at two critical points (ULS and TB) and several generic points in the dimer and Haldane phases of the BLBQ model. We find good qualitative (and quantitative) agreement with results obtained directly from DMRG. We briefly discuss the salient structural features of the entanglement spectrum in all the phases (but see ref.~\onlinecite{thomale2015entanglement} for a more detailed analysis). Extracting the central charges of the conformal field theories describing the two gapless critical points from numerical computation of the entanglement spectrum on finite open-boundary systems is made difficult by a slowly decaying oscillatory term in the entanglement entropy. However, we do obtain very good agreement for the central charge at the TB point, 1.505 as compared to the exact analytical value of $1.5$. At the ULS point, we obtain a larger discrepancy, c=$1.33$ compared to the analytical value of $2$.

We also introduce an algorithmic procedure that orthogonalizes an iMPS by breaking it down into its pure states. This is essential when dealing with degenerate ground states that appear upon Gutzwiller projection as is the case with points in the dimer phase of the BLBQ model. Having obtained the pure states, we can compute the entanglement spectrum for any state in the groundstate manifold. We check that the entanglement spectrum obtained from iMPS matches the one we obtain from the finite MPS procedure. Discrepancies naturally appear as we approach values close to the thresholds used to generate the finite MPS and infinite MPS.

The methods can be easily adapted to the study of systems on 2D ladders (infinite in length but with finite width). The iMPS unit cell is now formed by the tensors sitting on the width of the cylinder. We will explore the applications of Gutzwiller projected variational wavefunctions to the study of 2D quantum spin liquids in future publications. This method may also be applicable to topological states such as quantum Hall and fractional Chern insulators that are represented  as products of mean-field wavefunctions.

\begin{acknowledgments}

BKC acknowledges support from the Department of Energy grant DOE de-sc0020165.
This project is part of the Blue Waters sustained petascale computing project, which is supported by the National Science Foundation (awards OCI-0725070 and ACI-1238993) and the State of Illinois.
Blue Waters is a joint effort of the University of Illinois at Urbana-Champaign and its National Center for Supercomputing Applications. GKC was supported by the US National Science Foundation via grant no. 1839204. GKC also acknowledges support from the Simons Foundation via the Investigator Award and the Many-Electron Collaboration.

\end{acknowledgments}

\bibliographystyle{IEEEtran}
\bibliography{bib}
\clearpage
\appendix

\onecolumngrid

\renewcommand\thefigure{S\arabic{figure}}  
\renewcommand\thetable{S-\arabic{table}}

\counterwithout{equation}{section} 
\counterwithout*{equation}{section} 

\renewcommand{\thepage}{S\arabic{page}} 
\renewcommand{\thesection}{S\arabic{section}}

\setcounter{page}{1}
\setcounter{figure}{0}  
\setcounter{table}{0}
\setcounter{equation}{0}

\counterwithout{equation}{section} 
\renewcommand{\theequation}{S\arabic{equation}}

\section*{\Large{Supplementary Material}}
\setcounter{section}{0}

\section{Finite MPS boundary tensors}\label{supplement:boundary_tensors}
We can produce the left/right boundary tensors in either left/right canonical form.
Naturally, most of the time we will be concerned with producing the left boundary tensor in left canonical form and the right boundary tensor in right canonical form. Employing the notation used in the main text:

\begin{equation}
    A^{[1]\sigma_1}_{1,a_1} = \braket{\sigma_1|L^{1,N}_{a_1}}
\end{equation}

\begin{equation}
    A^{[N]\sigma_N}_{a_{N-1},1} = \braket{\sigma_{N-1}|R^{N-1,N}_{a_{N-1}}}
\end{equation}

Similar expressions exist for the left boundary tensor in right-canonical form and the right boundary tensor in left canonical form respectively:

\begin{equation}
    A^{[1]\sigma_1}_{1,a_1} = \braket{\sigma_1|\Lambda^{1,N-1}_{a_1}L^{1,N}_{a_1}}
\end{equation}

\begin{equation}
    A^{[N]\sigma_N}_{a_{N-1},1} = \braket{\sigma_{N-1}|\Lambda^{N-1,N}_{a_{N-1}}R^{N-1,N}_{a_{N-1}}}
\end{equation}

To produce a finite MPS in mixed-canonical form at site $l$:

\begin{equation}
\begin{split}
    A^{[1]\sigma_1}_{1,a_1} &= \braket{\sigma_1|L^{1,N}_{a_1}} \\
    A^{[i+1] \sigma_{i+1}}_{\alpha_{k}\alpha_{k+1}} &= \braket{ \sigma_{i+1} | \otimes  L^{i;N}_{\alpha_{k+1}} | L^{i+1;N}_{\alpha_k}} \quad \text{if} \qquad 1< i+1 \leq l \\
    A^{[i+1] \sigma_{i+1}}_{\alpha_{k}\alpha_{k+1}} &= \braket{ \sigma_{i+1} \otimes R^{i+1;N}_{\alpha_{k+1}} | R^{i;N}_{\alpha_k}} \quad \text{if} \qquad l< i+1< N \\
    A^{[N]\sigma_N}_{a_{N-1},1} &= \braket{\sigma_{N-1}|R^{N-1,N}_{a_{N-1}}}
\end{split}
\end{equation}

\section{iMPS Multi-site unit cell}\label{supplement:multi-site}
We can also consider iMPS with multiple sites unit cells - i.e. 

\begin{equation}
\ket{\Psi_{iMPS}} = \sum_{\sigma} \ldots \left( ABCD \right) \left( ABCD \right) \ldots
\end{equation} 

Our method will generate the uniform unit-cell tensor:

\begin{equation}
    T^{\sigma_{N+1}\sigma_{N+2}\sigma_{N+3}\sigma_{N+4}} = A^{\sigma_{N+1}}B^{\sigma_{N+2}}C^{\sigma_{N+3}}D^{\sigma_{N+4}}
\end{equation}

\begin{equation}
T^{\sigma_{N+1}\sigma_{N+2}\sigma_{N+3}\sigma_{N+4}}_{\alpha\beta} = \sum_\gamma C^{[2N+4]}_{\alpha\gamma} \left(\bra{ \sigma_{N+1} \sigma_{N+2} \sigma_{N+3} \sigma_{N+4}} \otimes \bra{ R^{N;2N}_{\beta}} \right) 
  \ket{R^{N,2N+4}_{\gamma}}
\end{equation}

We can obtain a more compact representation of $T$ by decimating the unit-cell tensor (i.e. inserting complete basis $|R^{N+1,2N+4}\rangle$,$|R^{N+2,2N+4}\rangle$, $|R^{N+1,2N+3}\rangle$ inside the inner product) and expressing it in terms of the following 4 on-site tensors:

\begin{equation}
    \tilde{A}^{\sigma_{N+1}}_{\alpha,j_1}  = \sum_{\gamma} C^{[2N+4]}_{\alpha\gamma} \left( \bra{\sigma_{N+1}} \otimes \bra{R^{N+1;2N+4}_{j_1}} \right) 
   \ket{R^{N,2N+4}}_{\gamma}
\end{equation}

\begin{equation}
    \tilde{B}^{\sigma_{N+2}}_{j_1j_2} = \left(  \bra{\sigma_{N+2}} \otimes  \bra{R^{N+2;2N+4}_{j_2}} \right) \ket{R^{N+1;2N+4}_{j_1}}
\end{equation}

\begin{equation}
    \tilde{C}^{\sigma_{N+2}}_{j_2j_3} = \left(  \bra{\sigma_{N+3}} | \otimes  \bra{R^{N+3;2N+4}_{j_3}} \right) \ket{R^{N+2;2N+4}_{j_2}}
\end{equation}

\begin{equation}
    \tilde{D}^{\sigma_{N+2}}_{j_3\beta} = \left(\bra{\sigma_{N+4}} \otimes  \bra{R^{N;2N}_{\beta}} \right) \ket{R^{N+3;2N+4}_{j_3}}
\end{equation}

So that:
\begin{equation}
    T^{\sigma_{N+1}\sigma_{N+2}\sigma_{N+3}\sigma_{N+4}} = \tilde{A}^{\sigma_{N+1}}\tilde{B}^{\sigma_{N+1}}\tilde{C}^{\sigma_{N+2}}\tilde{D}^{\sigma_{N+3}}
\end{equation}

Note that obtained in this way the tensors $\tilde{A}$,$\tilde{B}$,$\tilde{C}$ and $\tilde{D}$ are not necessarily uniform as they depend on the gauge choice of the Schmidt decompositions that generate $|R^{N+1,2N+4}\rangle$,$|R^{N+2,2N+4}\rangle$, $|R^{N+1,2N+3}\rangle$. 

\section{Schmidt decomposition of Slater Determinants and MPS tensor computation}\label{supplement:schmidt_tensor}

Consider an $N$-particle Slater determinant $|\Psi\rangle$ with support on an (indexed) $M$-site system. For a bipartition $[L]x[R] = [1,\ldots i]x[i+1,\ldots M]$, where $i$ is any site index $1 \leq i \leq M-1$, $|\Psi\rangle$ can be easily and efficiently brought to the following form \cite{peschel2012special}:

\begin{equation}
    |\Psi\rangle = \Pi^{N}_{k=1} \left[\sqrt{\epsilon_k} \phi^{\dagger}_{k;L} + \sqrt{1-\epsilon_k} \phi^{\dagger}_{k;R}\right]
\end{equation}

Here, $0 \leq \epsilon_k \leq 1$; $\phi^{\dagger}_{k;L}$ and $\phi^{\dagger}_{k;R}$ are operators that create orthogonal single orbital states with support in the left, respectively right, bipartitions.

The Schmidt decomposition of $|\Psi\rangle$ is obtained by expanding the above product:

\begin{equation}
    |\Psi\rangle = \sum_{\{l\},\{r\}} \lambda_{\{l_1,l_2,\ldots l_p;r_1,r_2,\ldots r_{N-p}\}} |\phi_{{l_1};L},\ldots.\phi_{{l_p};L}\rangle |\phi_{{r_1};R},\ldots.\phi_{{r_{N-p}};R}\rangle
\end{equation}

where 
$$
\lambda_{\{l_1,l_2,\ldots l_p;r_1,r_2,\ldots r_{N-p}\}} = \sqrt{\epsilon_{l_1}}\cdots \sqrt{\epsilon_{l_p}} \cdots \sqrt{1-\epsilon_{r_1}}\cdots \sqrt{1-\epsilon_{r_p}}
$$
are the Schmidt values and:

$$
|L_{\{l\}}\rangle = |\phi_{{l_1};L},\ldots.\phi_{{l_p};L}\rangle
$$

$$
|R_{\{r\}}\rangle=|\phi_{{r_1};R},\ldots.\phi_{{r_{N-p}};R}\rangle
$$

are the left/right Schmidt vectors. These are Slater determinants made up of subsets of the $N$-orbital sets $\{\phi_{;L}\}$ ($\{\phi_{;R}\}$).

The \textit{exact} Schmidt decomposition is given by an exponential number, $2^{N}$ of tuples $\left(\lambda, |L\rangle, |R\rangle \right)$. In practice, we will select only the tuples that have $\lambda$ greater than a threshold value. Due to the exponentially decaying nature of the Schmidt values, this will actually give us a very good approximation of the Slater determinant.

\subsection{Overlaps of Schmidt vectors}\label{supplement:overlaps}

Consider two right Schmidt vectors generated by $p$-orbital subsets $\{r^{1}\}$ and $\{r^{2}\}$.
Their overlap is given by:
\begin{equation}
    \langle R_{r^{1}} | R_{r^{2}}  \rangle = \langle \phi_{{r^1_1};R},\ldots.\phi_{{r^2_p};R}| \phi_{{r^2_1};R},\ldots.\phi_{{r^2_p};R}\rangle 
\end{equation}

So 
\begin{equation}
    \langle R_{r^{1}} | R_{r^{2}}  \rangle = \det{M_{\{r^1\}\{r^2\}}}
\end{equation}

\begin{equation}
    M_{\{r^1\}\{r^2\}} = 
    \begin{bmatrix}
   \braket{\phi_{{r^1_1};R}|\phi_{{r^2_1};R}} & \braket{\phi_{{r^1_1};R}|\phi_{{r^2_2};R}} &
   \hdots &
   \braket{\phi_{{r^1_1};R}|\phi_{{r^2_p};R}} \\
   \braket{\phi_{{r^1_2};R}|\phi_{{r^2_1};R}} & \braket{\phi_{{r^1_2};R}|\phi_{{r^2_2};R}} &
   \hdots &
   \braket{\phi_{{r^1_2};R}|\phi_{{r^2_p};R}} \\
   \vdots & \vdots & \ddots &\vdots \\
   \braket{\phi_{{r^1_p};R}|\phi_{{r^2_1};R}} & \braket{\phi_{{r^1_p};R}|\phi_{{r^2_2};R}} &
   \hdots &
   \braket{\phi_{{r^1_p};R}|\phi_{{r^2_p};R}}

   \end{bmatrix}
\end{equation}

Each of these matrices can be formed by selecting rows and columns from the matrix $O$ which needs to be computed only once.

\begin{equation}
    O = 
    \begin{bmatrix}
   \braket{\phi_{1;R}|\phi_{1;R}} & \braket{\phi_{1;R}|\phi_{2;R}} &
   \hdots &
   \braket{\phi_{1;R}|\phi_{N;R}} \\
   \braket{\phi_{2;R}|\phi_{1;R}} & \braket{\phi_{2;R}|\phi_{2;R}} &
   \hdots &
   \braket{\phi_{2;R}|\phi_{N;R}} \\
   \vdots & \vdots & \ddots &\vdots \\
   \braket{\phi_{N;R}|\phi_{1;R}} & \braket{\phi_{N;R}|\phi_{2;R}} &
   \hdots &
   \braket{\phi_{N;R}|\phi_{N;R}}

   \end{bmatrix}
\end{equation}

\section{IMPS with edge modes}\label{supplement:imps_edge}
Obtaining the iMPS involves producing mean-field ground states on two systems of different sizes. If the mean-field Hamiltonian has multiple zero energy orbitals, we need to make sure we form the two ground states using the ``same'' zero energy orbitals.
We next show how this works in the context of the SSH model.

$$
H_{SSH} = v \sum_n \left( c^{\dagger}_{n,1} c_{n,2} +\text{h.c.}\right)+ w\sum_n \left(c^{\dagger}_{n+1,1}c_{n,2} + \text{h.c.} \right)
$$

In the topological phase $v < w$, there will be two zero-energy modes (and hence 2-fold ground state degeneracy).

\begin{figure}[H]
  \centering
    \includegraphics[width=0.5\textwidth]{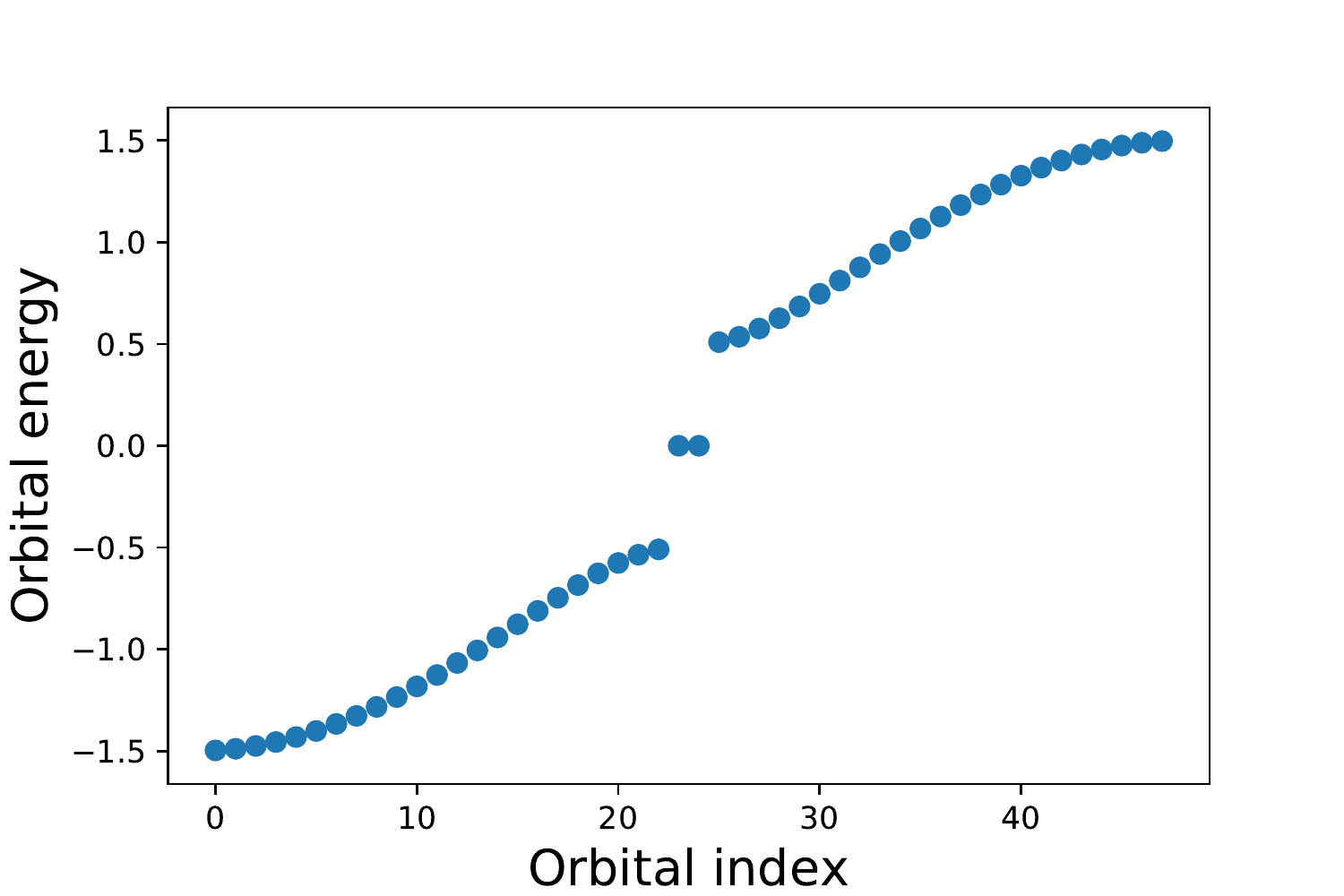}
      \caption{Single particle eigenvalues for the SSH model (eqn.~\ref{eqn:SSH}) for $N = 24$, $v=0.5$, $w = 1$.}

\end{figure}

\begin{figure}[H]
  \centering
    \includegraphics[width=0.5\textwidth]{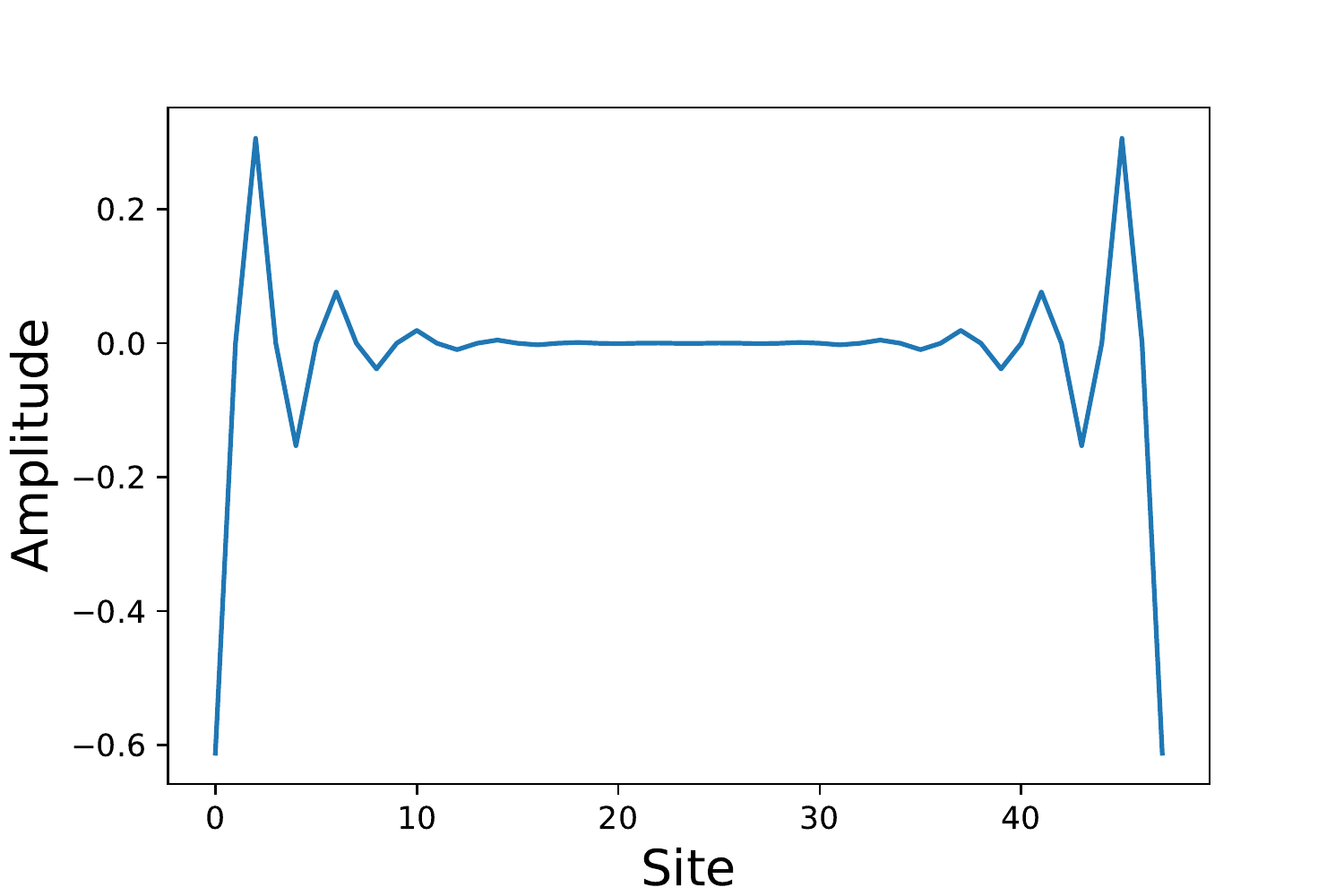}
  \caption{Edge mode eigenvector in real-space for $N = 24$, $v=0.5$, $w =1$ for the SSH model (eqn.~\ref{eqn:SSH}) .}

\end{figure}

Suppose the ground state produced in the first step of iMPS contains the zero-energy mode depicted in the figure. We picture the insertion step as introducing an unit cell in the middle of the chain. We generate the zero energy mode of the N+1=49 unit cell ground state by having the amplitudes on the first 24 unit cells and last 24 unit cells the same as the amplitudes of the zero energy mode on the N=48 unit cell system. In addition, the amplitudes on the 2 new sites introduced in the middle will be zero.

\section{Constant entanglement spectrum in the bulk for area-law entangled states}\label{supplement:constant_es}

We obtain the entanglement spectrum for the Slater Determinant ground state of the SSH model of eqn.~\ref{eqn:SSH} on a finite OBC chain with $N=24$ unit-cells, $v = 1$, $w=0.99$ and Schmidt value threshold $\epsilon = 10^{-12}$.

\begin{figure}[H]
    \includegraphics[width=0.5\textwidth]{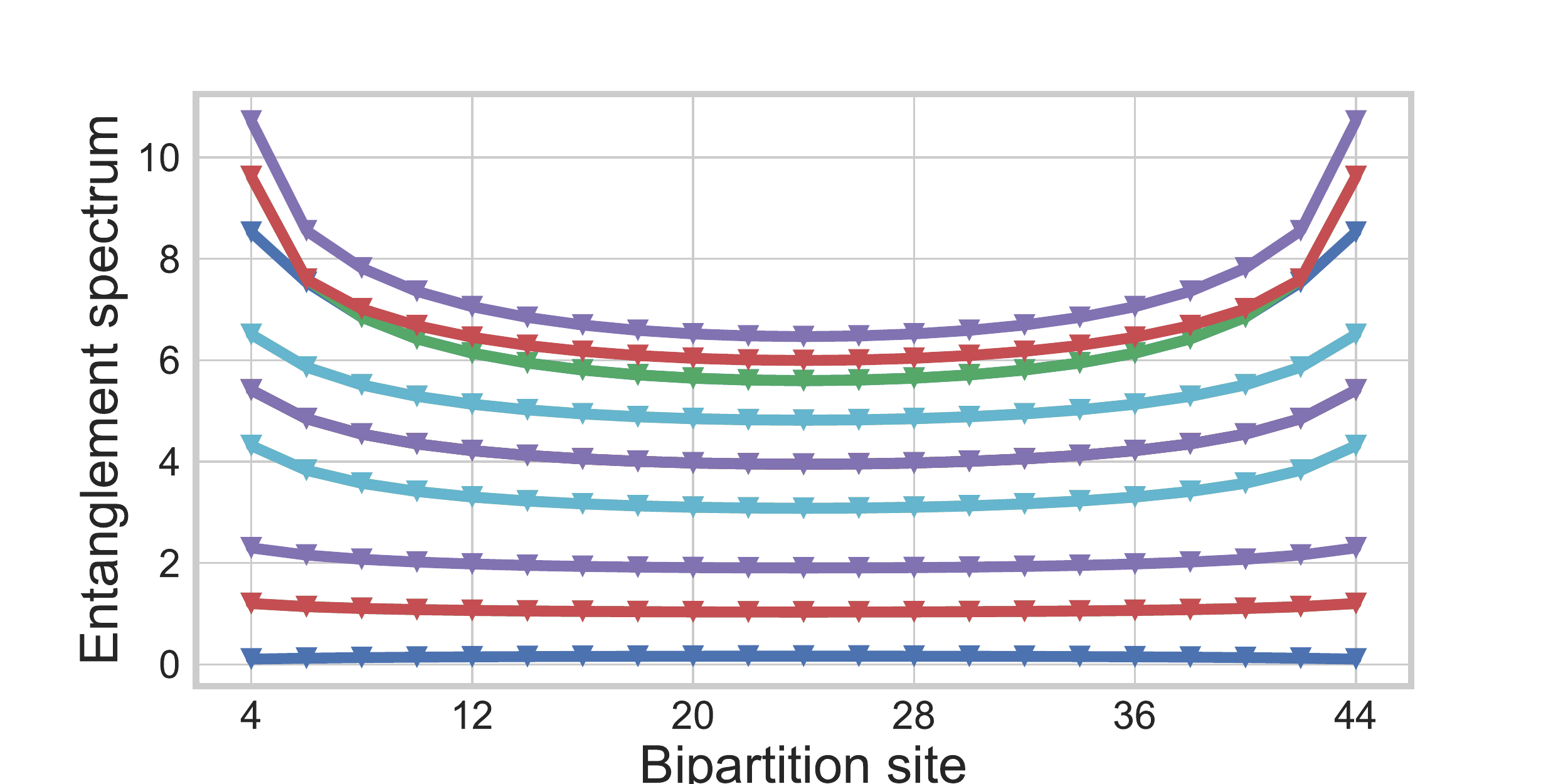}
    \includegraphics[width=0.5\textwidth]{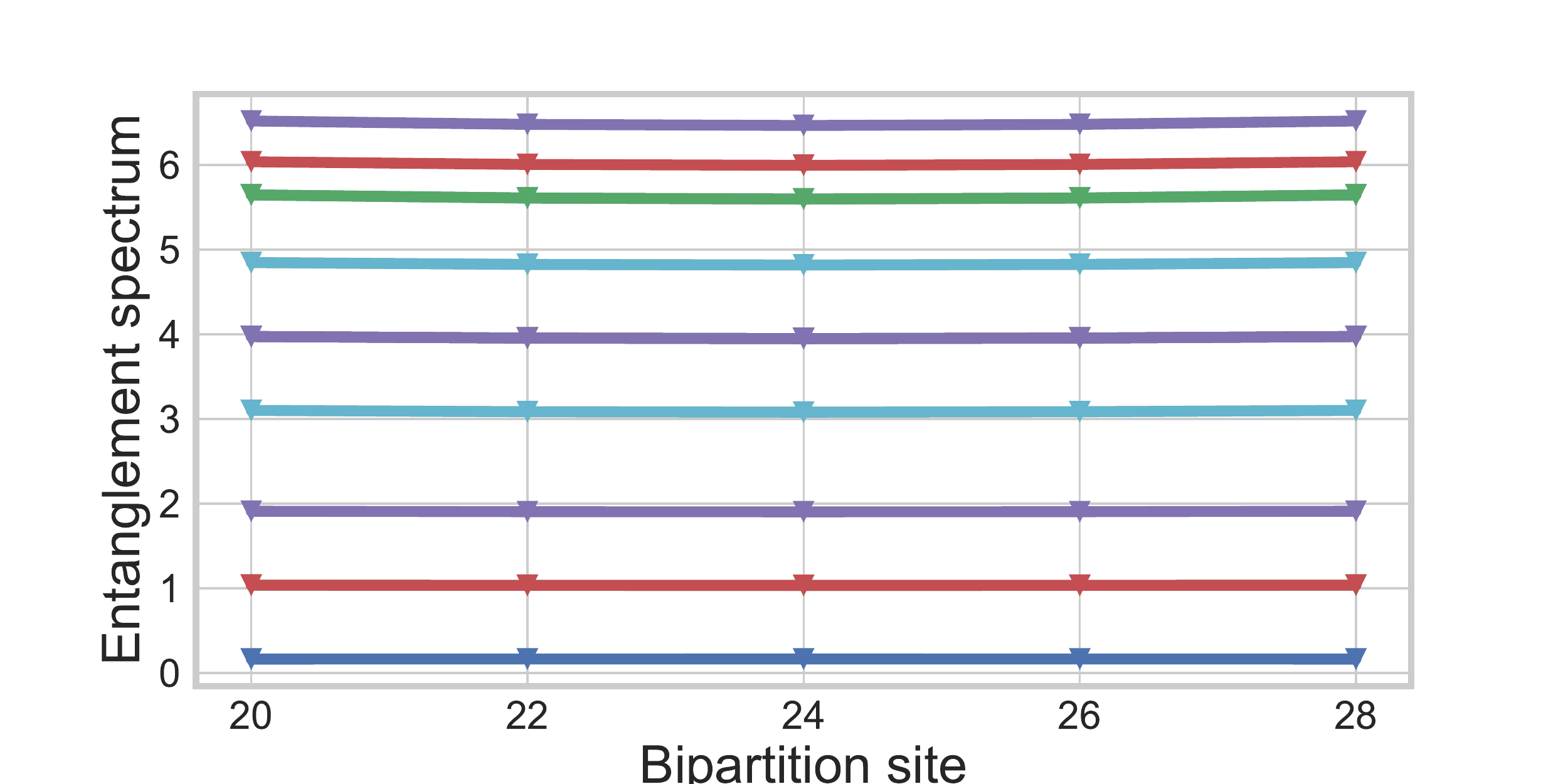}
  
  \caption{Top: Largest 16 entanglement levels (some levels are degenerate) for the SSH groundstate obtained on $N=24$ unit cells, $v=1$, $w = 0.99$. We are considering here the even cuts, i.e. cuts between unit cells (and not inside unit cells). Similar results hold for the odd cuts, i.e entanglement cuts inside the unit cells. Bottom: We show here the largest 16 entanglement levels (some levels are degenerate) for entanglement cuts around the middle of the chain (i.e ``bulk''). Note that the values have converged in bulk and hence our iMPS algorithm can be used.}
\label{fig:es_bulk}
\end{figure}

\section{iMPS orthogonalization procedure}\label{supplement:ortho_procedures}

In this subsection we present an intuitive derivation of the iMPS orthogonalization procedure given in ref.~\onlinecite{orus2008infinite}.

Upon constructing the left/right and uniform bulk tensors for our area-law entangled wavefunction, and performing a local operation (i.e Gutzwiller projection), we end up with non-orthonormal MPS of the following form (contraction of virtual indices implied):

\begin{equation}
    \ket{\Psi} = \sum_{\{\sigma\}} L^{\sigma_1}_1 L^{\sigma_2}_2 \ldots L^{\sigma_N}_N A^{\sigma_{N+1}} A^{\sigma_{N+2}} \ldots A^{\sigma_{N+2M}} R^{\sigma_{N+2M+1}}_N \ldots R^{\sigma_{2N+2M-1}}_2 R^{\sigma_{2N+M}}_1 \ket{\sigma_1 \ldots \sigma_{2N+2M}}
\end{equation}

We are now interested in putting this MPS into a mixed-canonical form with a cut between sites $N+M$ and $N+M+1$.

We rewrite the above MPS (and drop the physical indices for ease of presentation although they are implied) as:

\begin{equation} \label{eq1}
\begin{split}
\ket{\Psi} & = \sum \left(L_1 L_2 \ldots L_N A A \ldots A\right)_{1,\alpha} \left(A A \ldots A R_N \ldots R_2 R_1\right)_{\alpha,1} \\
 & =  \sum \ket{\Psi_L}_{\alpha} \ket{\Psi_R}_{\alpha}
\end{split}
\end{equation}

with $M$ uniform $A$ tensors inserted after $L_N$ and $M$ uniform $A$ tensors inserted before $R_N$;  $\ket{\Psi_L(R)}_{\alpha}$ are non-orthonormal vectors on the left(right) system bipartitions.

To orthogonalize the (say) left vectors we obtain their overlap matrix $M^L$ (which is hermitian). Upon diagonalization of $M^L$ we are able to obtain an orthogonal set of vectors on the left bipartition. We proceed in the same way for the set of right vectors.

\begin{equation}
M^L_{\alpha\beta} = _{\beta}\braket{\Psi_L|\Psi_L}_{\alpha}
\end{equation}
Since $M_L$ is hermitian, 
\begin{equation}
M_L = U_L D_L U^{\dagger}_L
\end{equation}

It then follows that
\begin{equation}
1 = D_L^{-1/2}U^{\dagger}_L \tilde{\ket{\psi_L}}^{\dagger}\tilde{\ket{\psi_L}} U_L D_L^{-1/2} 
\end{equation}
with $\tilde{\ket{\psi_L}}$ a vector formed from $\ket{\psi_L}_{\alpha}$ so that 

\begin{equation}
    \ket{\phi_L}_{\gamma} = \ket{\psi_L}_{1,\alpha} U_{L\alpha\gamma} D_L^{-1/2}
\end{equation} is now a set of orthonormal (Schmidt) vectors.

Similarly, for the right vectors we form the overlap matrix:

\begin{equation}
M^R_{\alpha\beta} = _{\beta}\braket{\psi_R|\psi_R}_{\alpha}
\end{equation}

\begin{equation}
(M^R_{\alpha\beta})^T = \sum R^{*} R^{*} \ldots R^{*}_N \ldots R^{*}_1 R^{T }_1 \ldots R^T_N R^T\ldots R^T
\end{equation}

Again $M_R$ is hermitian and can be diagonalized:
\begin{equation}
M^T_R = U_R D_R U^{\dagger}_R
\end{equation}

\begin{equation}
(M_R) = U^{*} D_R U^T_R
\end{equation}
and then
\begin{equation}
1 = D_R^{-1}U^{T}_R \left(\tilde{\ket{\psi_R}}\right)^{\dagger} \tilde{\ket{\psi_R}} U^*_R D_R^{-1/2}
\end{equation}
and so

\begin{equation}
\ket{\phi_R}_{\gamma,1} = (\ket{\psi_R}_{\alpha,1} U^*_{R\alpha\gamma} (D^{-1/2}_R)_{\gamma\gamma}
\end{equation}

\begin{equation}
\ket{\phi_R}_{\gamma,1} = {D_R^{-1/2}}_{\gamma\gamma}  {U^*_R}^T_{\gamma\alpha} \ket{\psi_R}_{\alpha,1}
\end{equation}

\begin{equation}
\ket{\phi_R}_{\gamma,1} = {D_R^{-1/2}}_{\gamma\gamma}  (U^{\dagger}_R)_{\gamma\alpha} \ket{\psi_R}_{\alpha,1}
\end{equation}

Going back to our original MPS representation and introducing the following resolutions of identity
\begin{equation}
I = U_L D_L^{-1/2} D^{1/2}_L U^{\dagger}_L
\end{equation}
and
\begin{equation}
I = U D_R^{1/2} D^{-1/2}_R U^{\dagger}
\end{equation}

we have:

\begin{equation} \label{eq1}
\begin{split}
\ket{\Psi}   & =  \sum \ket{\Psi_L}_{\alpha} \ket{\Psi_R}_{\alpha}
 \\ 
& = \sum \left(L_1 L_2 \ldots L_N A A \ldots A\right)_{1,\alpha} \left(A A \ldots A R_N \ldots R_2 R_1\right)_{\alpha,1} \\
& = \sum \left(L_1 L_2 \ldots L_N A A \ldots A\right)_{1,\alpha} \\ &\times 
\left(U_L D_L^{-1/2} D^{1/2}_L U^{\dagger}_L\right)_{\alpha,\alpha} \\ &\times 
\left(U D_R^{1/2} D^{-1/2}_R U^{\dagger}\right)_{\alpha,\alpha} \\ &\times 
\left(A A \ldots A R_N \ldots R_2 R_1\right)_{\alpha,1}
\end{split}
\end{equation}

We isolate the orthogonal vectors by collecting the terms:
\begin{equation}
\begin{split}
    \ket{\psi} &= \sum \left(\ket{\psi}_{\alpha}   
{U_L}_{\alpha\gamma} {D_L^{-1/2}}_{\gamma\gamma} \right) \\
&\times \left( D^{1/2}_L U^{\dagger}_L  U_R D_R^{1/2} \right)_{\gamma\delta} \\ 
&\times  \left({D_R^{-1/2}}_{\delta\delta} {U_R^{\dagger}}_{\delta\alpha} \ket{\psi_R}_{\alpha}\right)
\end{split}
\end{equation}

We thus obtain:

\begin{equation}
    \ket{\psi} = \ket{\phi_L}_{1,\gamma} \left( D^{1/2}_L U^{\dagger}_L  U_R D_R^{1/2} \right)_{\gamma\delta} \ket{\phi_R}_{\delta,1}
\end{equation}

By writing:
$
M_L = Y^{\dagger}Y
$
and
$
(M_R)^T = X X^{\dagger}
$
then 
$
YX = \left( D^{1/2}_L U^{\dagger}_L  U_R D_R^{1/2} \right)
$

and obtain the SVD of $YX=USV^{\dagger}$ to finally obtain the mixed canonical representation of the MPS:
\begin{equation}
    \ket{\psi} = \left( \ket{\phi_L} U \right) S \left(V^{\dagger}\ket{\phi_R} \right)
\end{equation}

The presence of the uniform tensor in the bulk significantly simplifies obtaining $M_L$ and $M_R$ in the $M \gg N$ limit, since the only terms that survives in the product of tensors solve the fixed point relations:
\begin{equation}
A^{\dagger}M_L A = \eta M_L
\end{equation}
where $|\eta|$ is the largest such value(s). Similarly for $M_R$
\begin{equation}
A^{*}M_R A^T = \eta M_R
\end{equation}
When viewed as vectors $M_L$ and $M_R$ solve the following eigenvalue equations
\begin{equation}
\left(\sum A \bigotimes A^* \right) M^T_R = \eta M^T_R 
\end{equation}

\begin{equation}
\left(\sum A^T \bigotimes A^{\dagger} \right) M_L = \eta M_L 
\end{equation}

which are the equations for finding the right(left) eigenvectors of the transfer matrix $\varepsilon = \left( \sum A \bigotimes A^* \right)$,
hence reproducing the imps orthogonalization method used in ref.~\onlinecite{orus2008infinite}.

\section{Other entanglement spectra figures for BLBQ dimer phase points}\label{supplement:symm_dimer}

\subsection{$(J,K)=(1,-2)$ entanglement spectrum comparison for odd cut}
\begin{figure}[H]
    \includegraphics[width=\columnwidth]{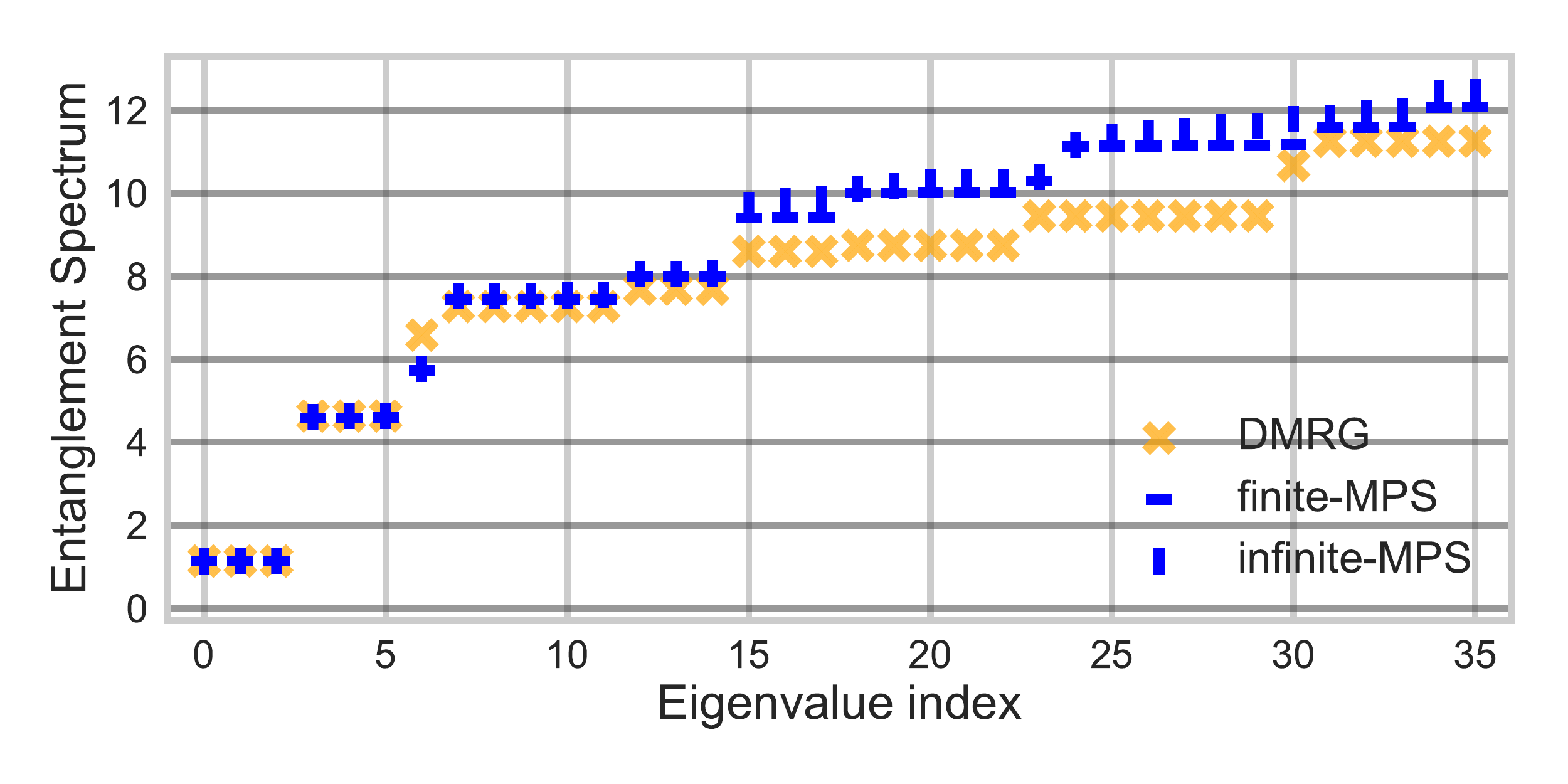}
    \caption{Entanglement spectra comparison (inside dimers cut) for $(J,K)=(1,-2)$ of the BLQB model (eqn.~\ref{eqn:blbq_ham}) variational point between fMPS, iMPS and DMRG.}
\label{fig:m1-m2-dmrg-fmps}
\end{figure}

\subsection{$(J,K)=(-1,-2)$ point}

\begin{figure}[H]
     \includegraphics[width=\columnwidth]{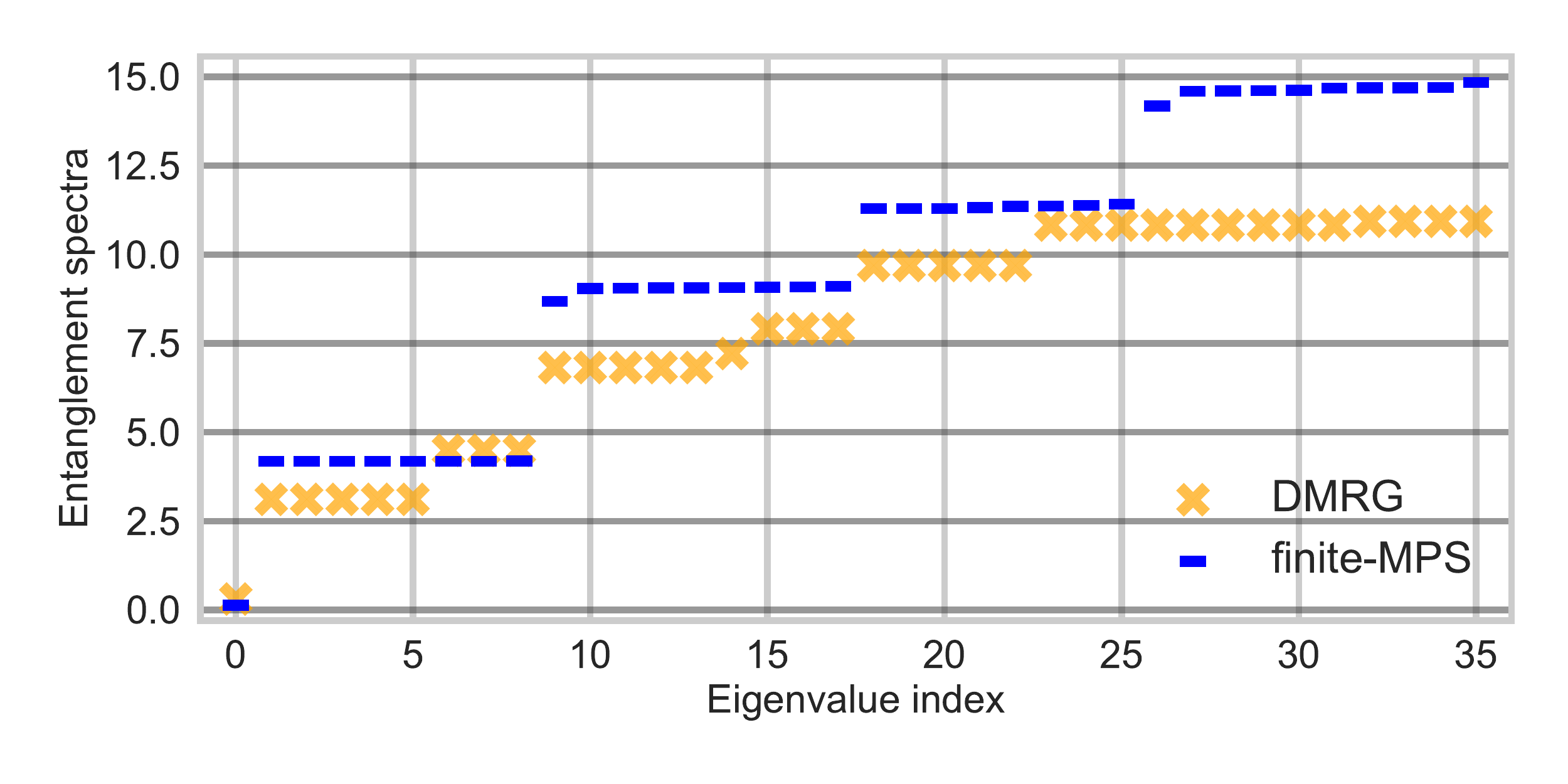}
    \includegraphics[width=\columnwidth]{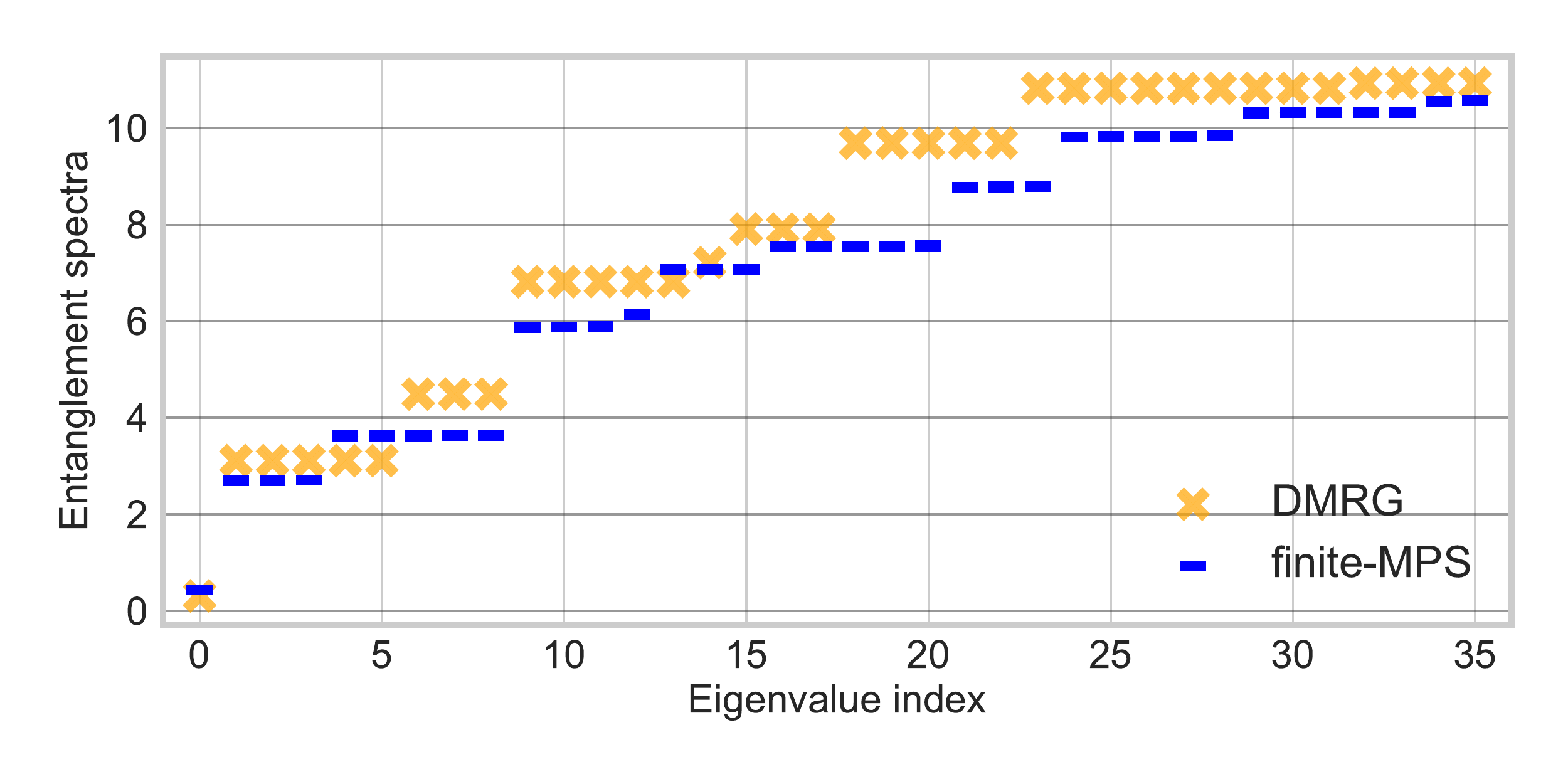}
    \caption{Comparison of entanglement spectra for the point $(J,K) = (-1,-2)$ of the BLBQ model (eqn.~\ref{eqn:blbq_ham}) between fMPS and DMRG. Top: Hopping parameter $\chi=0$ forces $SU(3)$ degeneracies in the entanglement spectrum. Bottom: Taking $\chi=0.1$, we see that the degeneracies of the entanglement spectrum are now broken to $SU(2)$.}
\label{fig:m1-m2-dmrg-fmps}
\end{figure}

\begin{figure}[H]
    \includegraphics[width=\columnwidth]{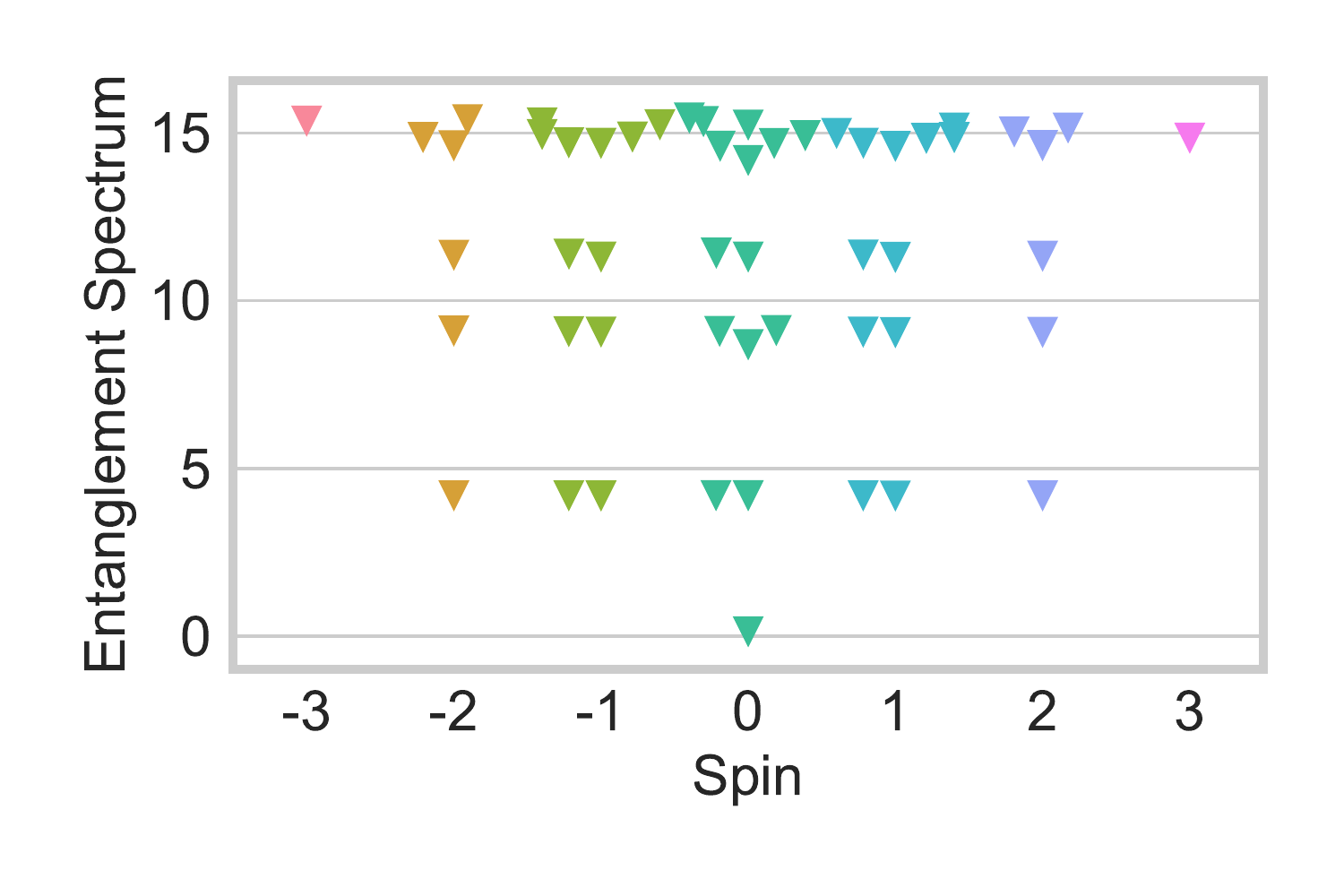}
    \includegraphics[width=\columnwidth]{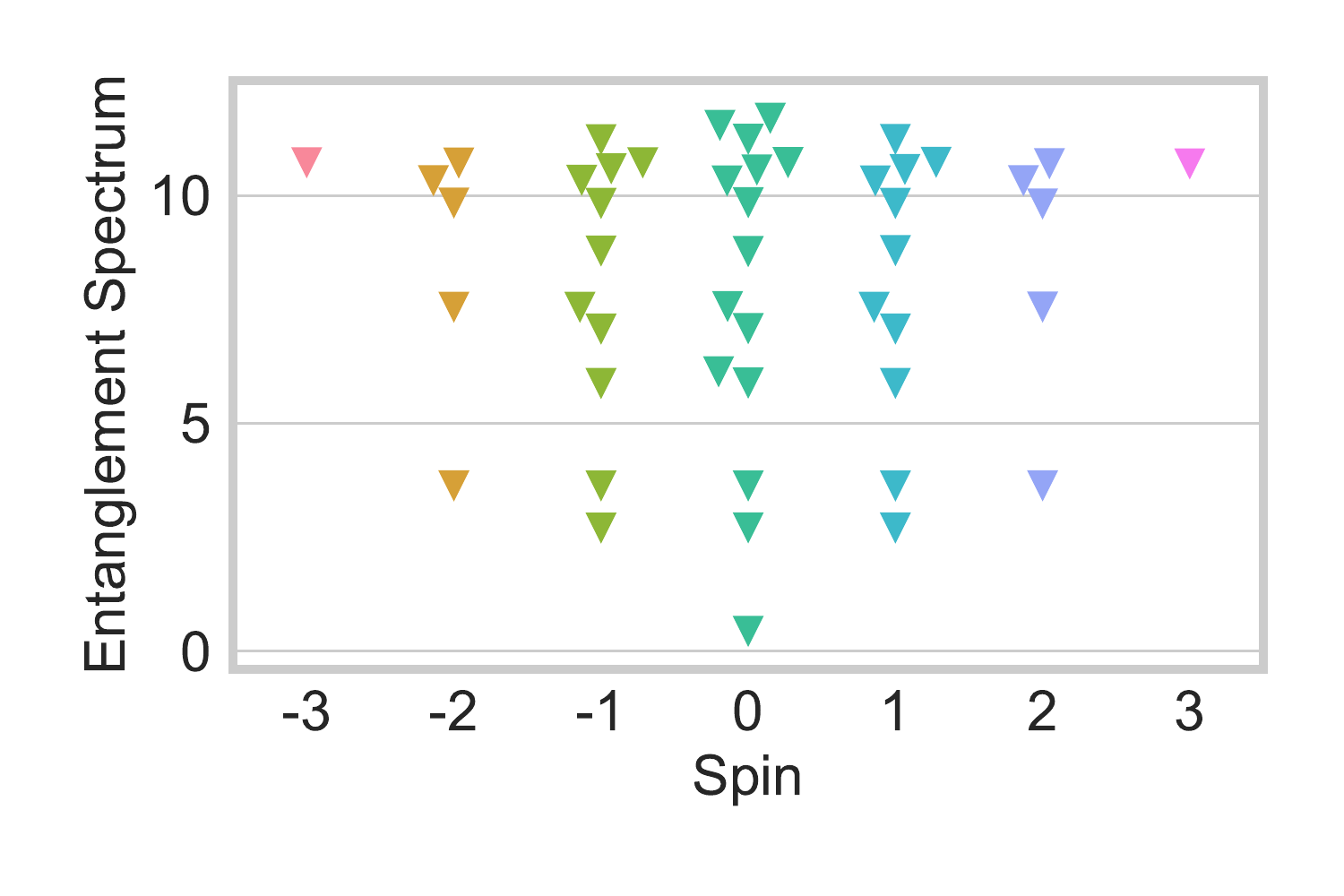}
    \caption{Spin resolved entanglement spectrum (between dimers cut) for the $(J,K)=(-1,-2)$ of the BLBQ model (eqn.~\ref{eqn:blbq_ham}) variational point obtained from fMPS. Top: hopping parameter $\chi = 0$. Bottom: hopping parameter $\chi = 0.1$}
\label{fig:dimer-symm}
\end{figure}

\section{iMPS tensors for spin $S=1/2$ MG chain}\label{supplement:mg_chain}

The Majumdar-Ghosh (MG) model is a simple exactly solvable 1D quantum spin chain. The MG Hamiltonian is:

\begin{equation}\label{eqn:ham_mg}
H_{MG} = J \sum_{n} S_n S_{n+1} + \frac{J}{2} \sum_n S_n S_{n+2}
\end{equation}

In the thermodynamic limit the groundstate is 2-fold degenerate. Two pure orthogonal states in the groundstate manifold are given by products of dimer singlets covering neighboring sites. In one state the dimers cover sites $(2n-1,2n)$; the other state is translationally shifted by one site and hence the dimers cover sites $(2n,2n+1)$ for integer $n$.

On a finite open-boundary conditions (OBC) $2N$ sites chain, we can write the even ground state as:

\begin{equation}\label{eqn:vbs_mg}
\ket{\psi_{e}} = \ket{S_{12}} \ket{S_{34}} \ldots \ket{S_{2N-1,2N}}
\end{equation}

where $\ket{S_{ij}} = \frac{1}{\sqrt{2}} \left( \uparrow_i \downarrow_j - \downarrow_i \uparrow_j \right)$ is the singlet configuration of 2 1/2 spins. 

The amplitude for a configuration $\ket{C_{\{\sigma\}}} = \ket{\sigma_1 \sigma_2 \ldots \sigma_{2N}}$ where $\sigma_i=\{\uparrow,\downarrow\}$ is:
\begin{equation}
    \braket{C|\psi_{e}} = \frac{1}{2^N} \times (-1)^{s_C}
\end{equation}

where $s_C$ counts the number of $\downarrow$ spins on odd sites.

We can obtain the groundstates of the MG chain within the Gutzwiller slave-fermion approach by using the following BdG slave-fermion mean-field Hamiltonian:

\begin{equation}
H_{BdG} = -\sum_{ij, \sigma}t_{ij}\left( c^{\dagger}_{i\sigma} c_{j\sigma} + \text{h.c.}\right) - 
\sum_{ij} \Delta_{ij} \left( c^{\dagger}_{i\uparrow} c^{\dagger}_{j\downarrow} + c^{\dagger}_{j\uparrow}c^{\dagger}_{i\downarrow} + \text{h.c.} \right) - 
\mu \sum_{i\sigma} c^{\dagger}_{i\sigma} c_{i\sigma} 
\label{eqn:bdg_mg}
\end{equation}

where we take $t_{n,n+1}=1$, $\Delta_{n,n+1}=-1$ and $\mu =-50000$.

\begin{figure}[h]
\includegraphics[width=0.5\linewidth]{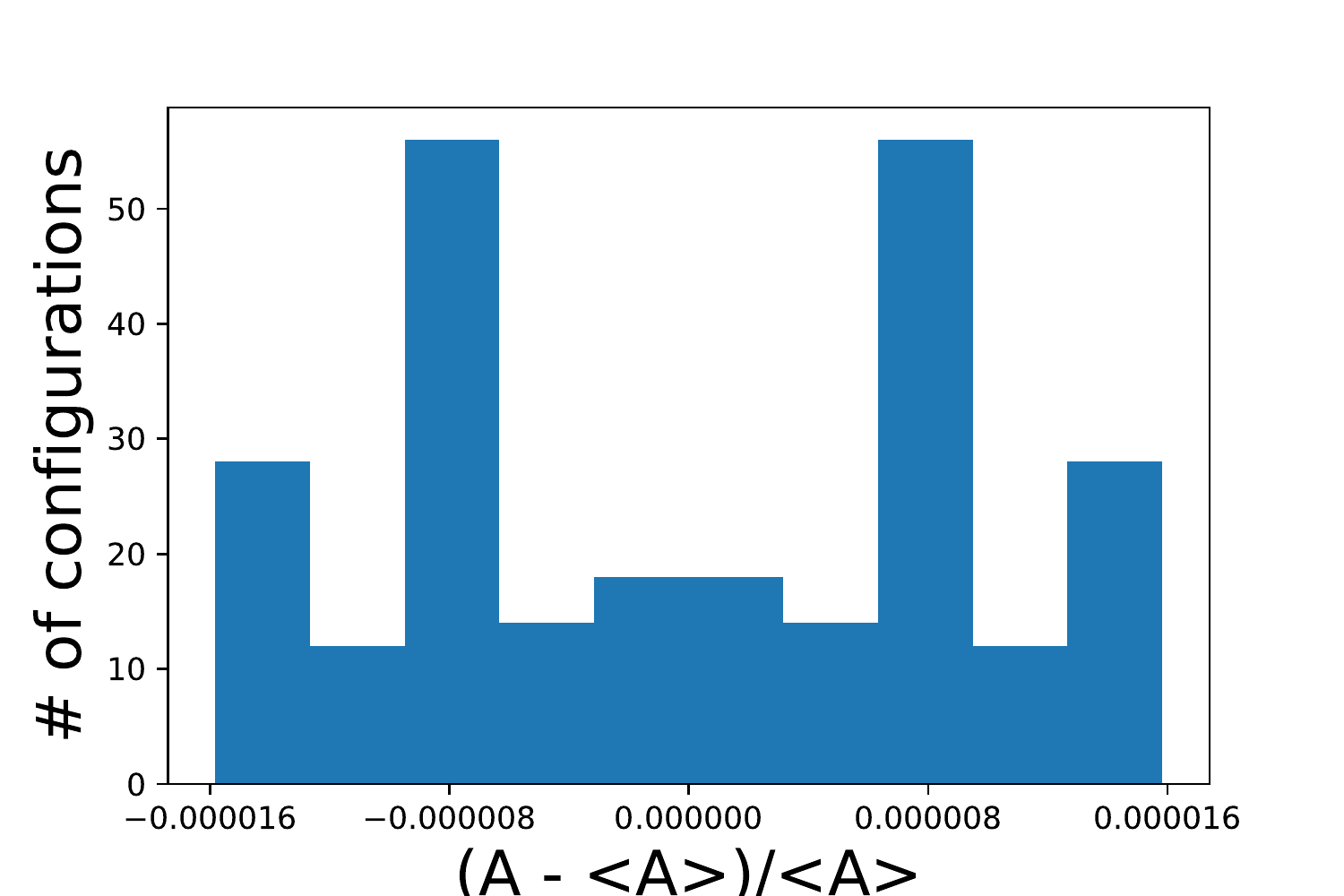} 
\caption{Histogram of the relative differences of amplitudes of the projected groundstate of eqn.~ \ref{eqn:bdg_mg} with $t_{n,n+1}=1$, $\Delta_{n,n+1}=-1$ and $\mu =-50000$  found for all configurations on $N=16$ sites.}
\label{fig:mg_amps_spread}
\end{figure}

First we produce the projected finite MPS on a small chain $N=16$ and check the amplitudes for all configurations have the proper magnitude and sign. We can see in figure \ref{fig:mg_amps_spread} that the relative error in the magnitude of amplitudes is at most $1.5\times 10^{-5}$.

We now obtain the iMPS representation using the procedures discussed in the main body of the text.

In particular we obtain the leading left/right eigenvectors of the 1-site transfer matrix and work with the positive left and positive right eigenvector and employ the standard iMPS orthogonalization. 
This produces the iMPS for the translationally invariant ($N \rightarrow \infty$) state $\ket{\psi_e} + T\ket{\psi_e}$ where $T$ translates the wave-function by one site .

The entanglement spectrum for this state is exactly given by: $[\log(2), \log(4), \log(4)]$. We report here the difference between the entanglement spectrum we obtain from iMPS and the true entanglement spectrum:

\begin{equation}
\delta_{\text{es}} = (6.59360277 \times 10^{-9}, -5.50256990 \times 10^{-6},  5.48941290\times 10^{-6})
\end{equation}

We also compute the energy in the thermodynamic limit. The ground state energy of the MG chain is exactly given by $-\frac{3}{8}$. We obtain the following:

\begin{equation}
    E_{MG} = -\frac{3}{8} - 3.438739293315507 \times 10^{-10}
\end{equation}

We obtain the left-canonical tensors in an arbitrary gauge. After a suitable rotation $A_L \leftarrow U^{\dagger}A_L U$, they are as follows:

\begin{equation}
    A^{\downarrow}_L = 
    \begin{pmatrix}
    0 & e^{i\phi_1} & 0 \\
    0 & 0 & 0 \\
    0.707103660 e^{-i\phi_2} & 0 & 0 
    \end{pmatrix}
    A^{\uparrow}_L = 
    \begin{pmatrix}
    0 & 0 & e^{i\phi_2}  \\
    -0.707104854 e^{-i\phi_1} & 0 & 0 \\
    0 & 0 & 0
    \end{pmatrix}
\end{equation}

Note that since the ground state is a singlet, there will be the same numbers of $\downarrow$ and $\uparrow$ tensors and hence the phases $\phi_1$ and $\phi_2$ will cancel out. We can further rotate the matrices by a unitary that changes the sign of the first column:
\begin{equation}
    U = \begin{pmatrix}
    i & 0 & 0 \\
    0 & 1 & 0 \\
    0 & 0 & 1
    \end{pmatrix}
\end{equation}

so that finally, we can put the 1-site left-canonical translationally invariant iMPS tensors in the form:

\begin{equation}
    A^{\downarrow}_L = 
    \begin{pmatrix}
    0 & 1 & 0 \\
    0 & 0 & 0 \\
    -1/\sqrt{2}+\epsilon_1  & 0 & 0 
    \end{pmatrix}
    A^{\uparrow}_L = 
    \begin{pmatrix}
    0 & 0 & 1  \\
    1/\sqrt{2} + \epsilon_2 & 0 & 0 \\
    0 & 0 & 0
    \end{pmatrix}
\end{equation}

where 
$
\epsilon_1 = -2.91581 \times 10^{-6}$
and 
$\epsilon_2 = 2.91918 \times 10^{-6}$. A further rotation brings the above tensors in their Krauss operator form \cite{ruiz2011tensor}.  To obtain the uniform tensors, notice that since
$\Lambda A_R = A_L \Lambda$, with $A_L$ and $A_R$ left and right canonical matrices respectively, we can obtain a ``uniform'' iMPS tensors
as 
$A = \sqrt{\Lambda}^{-1} A_L \sqrt{\Lambda} = \sqrt{\Lambda} A_R \sqrt{\Lambda}^{-1}$.
In this way we regain the MPS representation found in ref.~\onlinecite{karimipour2008matrix}:
\begin{equation}
\begin{split}
    A^{\downarrow}_L = 
    \begin{pmatrix}
    0 & 1 & 0 \\
    0 & 0 & 0 \\
    -1 + \epsilon_1  & 0 & 0 
    \end{pmatrix}\\
    A^{\uparrow}_L = 
    \begin{pmatrix}
    0 & 0 & 1  \\
    1 + \epsilon_2 & 0 & 0 \\
    0 & 0 & 0
    \end{pmatrix}
\end{split}
\end{equation}
and which by a further rotation can be brought to the form given in ref.~\onlinecite{perez2006matrix}.

Decomposing our iMPS into pure states, we can also obtain the $2$-site translational invariant tensors (see ref.~\onlinecite{perez2006matrix}, ref.~\onlinecite{ruiz2011tensor}). The 2-site unit cell uniform iMPS tensors for the even pure state (cuts between dimers) and odd pure state (cuts inside dimers) are given by:

\begin{equation}
\begin{split}
    A^{\sigma\sigma'}_{\text{even}} = B^{\sigma}C^{\sigma'} \\
    A^{\sigma\sigma'}_{\text{odd}} = C^{\sigma}B^{\sigma'}
\end{split}
\end{equation}
with the $B$ and $C$ tensors numerically obtained as:

\begin{equation}
    B^{\downarrow} = 
    \begin{pmatrix}
     1 + \epsilon_1 & 0
    \end{pmatrix}
    \qquad
    B^{\uparrow} = 
    \begin{pmatrix}
     1  & 0
    \end{pmatrix}
\end{equation}

\begin{equation}
    C^{\downarrow} = 
    \begin{pmatrix}
    0 \\
    1
    \end{pmatrix}
    \qquad
    C^{\uparrow} = 
    \begin{pmatrix}
     -1 + \epsilon_2 \\
     0
    \end{pmatrix}
\end{equation}
where $\epsilon_1 = -7.1400000000165775 \times 10^{6}$ and  $\epsilon_2 = 1.09999999997612 \times 10^{-6}$. The bond dimension for the 1-site translationally invariant iMPS representation is naturally larger. 

\end{document}